\RequirePackage{fix-cm}
\documentclass{svjour3}                     % onecolumn (standard format)
\smartqed  % flush right qed marks, e.g. at end of proof
\usepackage{graphicx}
\usepackage{mathptmx}      % use Times fonts if available on your TeX system
\usepackage{latexsym}
\usepackage{algorithmic}
\usepackage{textcomp}
\usepackage{booktabs}
\usepackage{tabularray}
\usepackage[flushleft]{threeparttable}
\usepackage{multirow}
\usepackage{tcolorbox}
\usepackage{stfloats}
\usepackage{soul}
\usepackage{enumitem}
\usepackage[T1]{fontenc}
\usepackage[utf8]{inputenc}
\usepackage{newtxtext,newtxmath}
\usepackage{xspace}
\usepackage{comment}
\usepackage[normalem]{ulem}
\usepackage[colorlinks=true,urlcolor=blue,citecolor=blue,linkcolor=blue,bookmarks=true,breaklinks=true]{hyperref}
\usepackage[nobiblatex]{xurl}
\usepackage[style=authoryear,natbib=true]{biblatex}
\usepackage{longtable}
\usepackage{xltabular}
\usepackage{CJKutf8}
\usepackage{listings}
\usepackage{color}
\definecolor{dkgreen}{rgb}{0,0.6,0}
\definecolor{gray}{rgb}{0.5,0.5,0.5}
\definecolor{mauve}{rgb}{0.58,0,0.82}
\definecolor{codegreen}{rgb}{0,0.6,0}
\definecolor{codegray}{rgb}{0.5,0.5,0.5}
\definecolor{codepurple}{rgb}{0.58,0,0.82}
\definecolor{backcolour}{rgb}{0.95,0.95,0.92}

\usepackage[toc]{appendix}
\usepackage[section]{placeins}
\pdfoutput=1
\DeclareMathAlphabet{\mathbbmsl}{U}{bbm}{m}{sl}
\newcommand{\bihui}[1]{\textcolor{black}{ #1}}
\newcommand{\bj}[1]{\textcolor{black}{ #1}}

\journalname{Empir Software Eng}
\begin{document}

\title{Impact of Extensions on Browser Performance: An Empirical Study on Google Chrome}
% \subtitle{Do you have a subtitle?\\ If so, write it here}


% \author{\fnm{Bihui} \sur{Jin}   \and
%         \fnm{Heng} \sur{Li}     \and
%         \fnm{Ying} \sur{Zou}
% }

\author{Bihui Jin   \and
        Heng Li     \and
        Ying Zou
}


% \institute{\textbf{Bihui Jin}, \textbf{Ying Zou}\at \orgdiv{Department of Electrical and Computer Engineering} \at
%               \orgname{Queen's University}, \orgaddress{\city{Kingston}, \state{ON}, \country{Canada}} \at
%               E-mail: \{b.jin, ying.zou\}@queensu.ca
%            \and
%            \textbf{Heng Li} \at
%            \orgdiv{Department of Computer Engineering and Software Engineering} \at
%            \orgname{Polytechnique Montréal}, \orgaddress{\city{Montréal}, \state{QC}, \country{Canada}} \at
%             E-mail: heng.li@polymtl.ca
% }

\institute{\textbf{Bihui Jin}, \textbf{Ying Zou}\at Department of Electrical and Computer Engineering \at
              Queen's University, Kingston, ON, Canada \at
              E-mail: \{b.jin, ying.zou\}@queensu.ca
           \and
           \textbf{Heng Li} \at
           Department of Computer Engineering and Software Engineering \at
           Polytechnique Montréal, Montréal, QC, Canada \at
            E-mail: heng.li@polymtl.ca
}

\date{Received: date / Accepted: date}

\maketitle

\begin{abstract}
    Web browsers have been used widely by users to conduct various online activities, such as information seeking or online shopping.
 To improve user experience and extend the functionality of browsers, practitioners provide mechanisms to allow users to install third-party-provided plugins (i.e., extensions) on their browsers. 
However, little is known about the performance implications caused by such extensions.
%In this paper, we conduct an empirical study to understand the performance impact that extensions have on Google Chrome, the most popular browser. 
In this paper, we conduct an empirical study to understand the impact of extensions on the user-perceived performance
% \heng{let's use user-perceived performance to motivate why we consider energy consumption and page load time} 
(i.e., energy consumption and page load time) of Google Chrome, the most popular browser. 
% We cluster the total of 61 extensions available in 12,423 to select a representative set of 72 extensions from 11 categories 
We study a total of 72 representative extensions from 11 categories (e.g., Developer Tools and Sports).
% \bihui{R3.1, 3.7 - In this work, we algorithmically select a total of 72 extensions, consisting of 11 extensions that do not use privacy practices chosen based on popularity, and 61 extensions selected through clustering.}
% We design experiments to study the impact of 72 extensions across 11 categories on the performance of Google Chrome, the most popular browser, in terms of the energy consumption and page load time of the browser.
We observe that browser performance can be negatively impacted by the use of extensions, even when the extensions are used in unintended circumstances (e.g., when logging into an extension is not granted but required, or when an extension is not used for designated websites). %\bj{not log in to the extension or not grant the extension permission to access the webpage})
%or are not active (e.g., not active for designated websites or fully deactivated).
We also identify a set of factors that significantly influence the performance impact of extensions, such as code complexity and privacy practices (i.e., collection of user data) adopted by the extensions.
% \heng{make sure this is updated \bihui{- was updated}} % are significantly correlated with its performance impact, as certain privacy practices may require additional resources and permissions. 
%We also observe that, surprisingly, extensions with larger code sizes tend to be more energy efficient (e.g., due to encompassments of comprehensive and optimized functionalities within a single package).
%From the results of our empirical study, we observe that (i) extensions in different activation modes (e.g., not logged into the extension) negatively affect the page load time by up to 80.27\%, the page load energy consumption by up to 15.57\%, and the stabilized energy consumption by up to 8.79\%, compared to the performance without the extensions used; (ii) the larger the size of an extension installation package or the more popular the extension, the lower the energy consumption; and (iii) the activation mode of extensions and the use of privacy practices in the extension significantly impact on the overall browser performance.
%Our observations help raise awareness of the performance impact of using the extensions and aid practitioners in further optimizing the development of browser extensions.
Based on our empirical observations, we provide recommendations for developers and users to mitigate the performance impact of browser extensions, such as conducting performance testing and optimization for unintended usage scenarios of extensions, or adhering to proper usage practices of extensions (e.g., logging into an extension when required).

\end{abstract}

\section{Introduction}
Online services, such as information seeking, video streaming, or social networking services, rely on web browsers  as the user interfaces to allow the users to interact with the provided services. 
In particular, Internet video dominates global Internet traffic, accounting for 71\%, and the web/data category (e.g., media and entertainment services and banking applications) constitutes 12\% of all Internet traffic in 2022 \citep{engTrace}.
With the growth of Information and Communication Technology (ICT), the median size of desktop web pages has surged by 336\% in a decade, escalating from 468 KB in 2010 to 2042 KB in 2020 \citep{engTrace}.
When a web browser takes longer to load a webpage or runs slower, it can be frustrating for users and leads to diminished productivity or even customer attrition~\citep{Tian, Borgolte, 10.1145/3341617.3326142}. 

Energy usage is considered as a non-trivial quality attribute of software products \citep{Pang}. 
Over the last three decades, the total energy consumption of ICT has surged by a staggering  822.79\%, increasing from 2182.72 TWh per year in 2001 to 17959.11 TWh per year in 2030~\citep{Trends2030}. 
Energy consumption is a concern for browser users, particularly for users of battery-powered devices (e.g., laptops)~\citep{Banerjee, Kor}. 
Web browsers operating on portable devices, such as laptops and mobile phones, drain a substantial amount of power to keep the webpage alive, update the content of webpages, or retain multiple tabs~\citep{10.1145/3417113.3423000}.
Moreover, the excessive energy consumption of electricity-power devices (e.g., desktops) has significant environmental impact \citep{Amsel, San}.
% \bj{We focus on the performance properties that directly impact user experience.} 
It is preferable that the software can be energy efficient, extend battery life, and enhance the overall user experience~\citep{Pang}. 
The current studies  \citep{Tiwari, Amsel, Fei} on energy consumption tend to predominately focus on CPU usage while neglecting the actual patterns of energy consumption resulting from software.

\bj{The performance of web browsers (e.g., energy consumption or page load time) plays a crucial role in shaping user experience and ensuring sustainability.}
%A payload is the period from the time a webpage starts loading until it finishes loading. 
%Payload performance includes the energy and time consumed while loading a page, waiting for data transmission and server response and rendering the webpage.
%The stabilized performance refers to the energy consumption of the CPU and memory after a webpage finishes loading.
% This metric differs from payload performance, which focuses on the
% energy consumption during the loading process. 
Web browsers, such as Google Chrome, typically support a variety of extensions that allow users to extend the browsers' functionalities, %substantial number of extensions that dynamically analyze code, send requests to servers, and use I/O operations. 
such as advertisement (ad) blocking, password management, and language translation.
However, the extensions can consume additional resources, such as processing power, memory, and network bandwidth, and may potentially affect the performance of browsers. %payload and stabilized performance of the other applications running on the same hardware environment. 
\bj{Prior work \citep{Pearce,Merzdovnik,Borgolte} has studied extensively on browser performance. 
Nonetheless, the prior work  has primarily examined the impact of browser extensions on either page load time or energy consumption of particular types of extensions.}
% There is a limited number of studies on the impact of the extensions on the performance of browsers.
% Therefore, studying the energy consumption and runtime performance of extensions is critical to understand the impact of extensions on computer performance.
% Other existing work \citep{Pearce,Merzdovnik,Borgolte} primarily focuses on particular types of extensions and \bj{to assess the impact of browser extensions on either page load time  or energy consumption.}
For example, \citet{Pearce} 
% is close to this work, which 
studies the energy effect of three ad blocking extensions in the Chrome browser during page loading and finds that open-source ad blockers reduce the waiting time for ads to load and decrease power consumption. % to run the ads on the computer. 
% However, \citet{Pearce} investigates only the ad blocking extensions.
% Furthermore, there is no evidence and similar studies indicating whether the computer performance (i.e., 
%without \bihui{actually monitoring the energy usage and considering energy impact on} the browser.
% are varied by using various functional types of extensions. 
% Little is known about the impact of plugins on energy consumption and runtime performance. 
\citet{Borgolte} study the impact of eight privacy-focused extensions on browser performance in terms of page load time and CPU time. 
% They observe that a browser with privacy-focused extensions performs similarly or better than a browser without extensions. 
Thus, the results of aforementioned studies may not be generalizable to other categories of extensions.

% \heng{We have not mentioned anything about websites yet. I suggest we only keep ``wider variety of extension types'' here, and mention that we also study how the configurations and other factors impact browser performance. Move the explanation for the number of websites to the related work section (when discussing the difference) or threats to validity. \bihui{-moved to the related work section}}
% \bihui{R2.5- In contrast, our work encompasses a wider variety of extension types but focuses on a smaller set of websites, which is a quid pro quo trade for our approach. Nevertheless, it is important to note that the websites used in our experiments are highly representative as they are among the most commonly utilized by users worldwide.  Therefore, despite the reduced number of websites, our findings still hold relevance and provide valuable insights into the performance of different extension types across widely-used online platforms.}
\bj{To address the limitation of the existing work, we offer a comprehensive approach to understand the impact of the extensions on the user-perceived performance of the browser. 
More specifically, our approach consists of the following aspects: }

\begin{itemize}[leftmargin=*,topsep=0pt]
\item We encompass a wider variety of extension types by selecting 72 representative extensions from 11 categories (e.g., shopping, blogging, or accessibility).

\item We study  performance metrics, including both page load time and energy consumption, as energy consumption and page load time directly impact user experience~\citep{Palomba, Chan-Jong-Chu, Janssen, Tian, Borgolte, Hindle}. 

\item We investigate energy consumption by breaking  down energy consumption from two phases, namely, page load and stabilized energy consumption, for a detailed analysis. 

\item We systematically examine how the configurations (e.g., active or inactive) and influential factors (e.g., code metrics) of extensions impact browser performance.
\end{itemize}

We organize our paper along  three research questions (RQs) \bihui{in a progressive manner. 
Specifically, RQ1 investigates the overall performance impact of browser extensions.
RQ2 further studies the difference brought by various activation modes of the extensions, and RQ3 systematically examines a wide range of factors of extensions to identify the factors that can significantly impact browser performance.}

% We use Selenium\footnote{\url{https://www.selenium.dev}} to collect 110,240 extension records from 11 categories of Chrome Web Store\footnote{\url{https://chrome.google.com/webstore}} which is open to Chrome users.
% We use clustering algorithm, K-medoids \cite{kmedoids1,kmedoids2}, to cluster the feature based on the use of extension privacy practices (e.g., location, web history, user activity, web-site content, etc.). 
% Eleven extensions, one from each category, closest to each cluster center are selected from each of the five clusters. Additionally, six extensions are selected to replace the six inactive extensions. As a result, 61 representative extensions are selected for inclusion in our study.
% Test scenarios are then customized and generalized for the selected 61 representatives to imitate users' behavior. By monitoring the response time of payload (i.e., page loading time), energy consumption of payload (i.e., energy consumption while loading the page), and the energy consumption after the page completes loading, i.e., stabilized energy consumption, we answer the following three research questions to understand the heuristics of extensions.

 \begin{description}[style=unboxed,leftmargin=0cm,topsep=0pt]
    \item \textbf{RQ1: How do extensions impact browser performance?}
%Extensions can consume energy from the computer in various ways, such as CPU computation and memory usage. 
Practitioners may not be aware that extensions can impair the performance of browsers and degrade the user experience. %other applications running on the same computer. We observe that when a user uses the extension in a particular way (i.e., logging in the extension, granting the page access, and running on the target webpage), the overall page load time, the page load energy consumption, and the stabilized energy consumption increase by 1.00\%, 15.57\%, and 1.44\%, respectively.
We analyze the impact of extensions on the browser performance when they are used in the expected mode (i.e., fully-loaded)
and observe that the use of extensions can lead to a statistically significant impact on the browser performance, with the largest negative impact on the load time energy consumption. 
%Our observations suggest that browser and extension developers should pay attention to the performance impact of browser extensions, particularly in regard to energy efficiency.

    \item \textbf{RQ2: How do the activation modes of extensions affect browser performance?}
    Users do not always interact with extensions in the intended or desired manner (i.e., activation modes of extensions).
    \bihui{The activation modes of an extension refer to the different ways users may use an extension. 
    For example, when using extensions, a user has the options to use the extension after login or use the extension without login. 
    Logged or not logged into the extension are two activation modes.}
    Extensions may have different performance impacts depending on different activation modes.
    %We find that different activation modes of extensions result in changes in browser performance. 
    By studying how different activation modes of extensions impact the browser performance, 
    we find that browser performance can be negatively impacted by extensions even when they are used in unexpected circumstances or are not active (e.g., not used for designated websites).
    Such unintended usage scenarios may even lead to a worse performance impact than the fully-loaded mode, suggesting the need for performance testing and optimization for such scenarios.
    %The improper use of the extension (e.g., ignoring requests for login and page access permission) negatively affects the performance the most. 
    %The activation modes of extensions can have a significant impact on performance, 
    % the performance of response time of payload by up to 80.27\%, of energy consumption of payload by up to 14.82\%, and of stabilized energy consumption by up to 8.79\%. 
    %leading to a statistically significant increase in the page load time by 80.27\%, the page load energy consumption by 14.82\%, and the stabilized energy consumption by 8.79\%.
    % \textbf{RQ2-2:} How are energy consumption and run-time performance correlated?
%Extension users should be aware of such performance impact to optimize their configurations of the extensions. Our findings also suggest that browser developers should take action to reduce the performance impact of extensions when they are not providing proper functionalities (e.g., when in an inactive mode) to users.

    \item \textbf{RQ3: What factors of extensions influence browser performance?} 
    When developers create extensions, they need to consider various factors that may potentially affect browser performance. 
    % \bihui{R1.9 - the sizes of various file types in extensions and code metrics (e.g., Lines of Code) related to the source code of the extensions}.
    % \heng{Maybe better use examples that developers could choose during the design (e.g., collection of user data) \bihui{-revised}}
    % the use of privacy practices () to track location and user activities, to analyze web histories, or to monitor personal communications. 
    In this RQ, we conduct quantitative evaluations on the factors of extensions that could impact browser performance.
    For instance, we observe that the adopted privacy practices of extensions (e.g., \textit{personal communication} and \textit{website contents}) can significantly impact browser performance, suggesting the need to consider the performance impact of privacy practices when developing and using extensions.
    \bj{Elevated code complexity (e.g.,  the number of functions) can also contribute to heightened energy consumption.
    Furthermore, extension developers ought to be mindful that certain file types (e.g., SVG image files and OGG audio files) used in extensions adversely affect browser performance.}
    % Developers create extensions using various decisions that may potentially affect browser performance, such as privacy practices (e.g., tracking location, and user activity). %, to fulfill the development demands.
    % By investigating \bihui{R3.5 - the impact of a comprehensive set of factors quantitatively,}
    % the relationship between the various factors of extensions and browser performance, 
    %We observe that extensions with larger code sizes tend to be more energy efficient.
    %Besides, the use of privacy practices of extensions has a significant impact on browser performance. 
    %Our findings provide insights for users to select performance efficient extensions and for developers to optimize their development decisions.
    %the usage of privacy practices in an extension and its work environment are two factors negatively impacting the browser performance. Additionally, we find that prevalent extensions (i.e., having a high rating score) or extensions with a larger installation package size tend to be more energy efficient.

\end{description}

% \heng{Moved to be right before the RQs}
 % \bihui{R3.1 - RQ1 investigates the impact of extensions on browser performance in a theoretical activation mode, RQ2 explores uncertainties in different activation modes, and RQ3 scientifically confirms the influence of activation modes on browser performance while identifying significant factors.}
In summary, the main contributions of this paper are three-fold:
 
\begin{itemize}[leftmargin=*,topsep=0pt]
    
    %\item Propose an approach to categorize the extensions based on the usage of privacy practices and the work environments of the extensions.

    \item We provide empirical evidence of the performance impact of browser extensions and their activation modes.
    
    %\item Investigate the impact of extensions on browser performance.
    
    %\item Evaluate the impact of the activation modes of extensions on browser performance.
    %Determine statistically the degree of changes in browser performance regarding the activation modes of extensions.

    \item We identify the factors that contribute to the performance impact of browser extensions.

    \item We provide recommendations for developers to conduct performing performance testing and to optimize  unintended usage scenarios of extensions, as well as suggest users to adhere to proper usage practices in order to mitigate the performance impact of extensions. 
    
    %\item Identify the factors that contribute to the energy consumption and the page load time.

    %\item Share our replication package~\cite{RepPackage} for future work to build on our work.
    % \item An empirical study arouses awareness of researchers and practitioners about the impact of computer performance caused by the use of extensions in Google Chrome.
\end{itemize}

\textbf{Paper organization.} 
% We provide the background on extensions in Section \ref{BC}. 
We present the design of experiments in Section \ref{ED}. We discuss our approaches and results in Section \ref{ER}. %In Section \ref{IMP}, we discuss the implications of this study. 
In Section \ref{TV}, we describe the threats to validity. We give an overview of related works in Section \ref{RW}. Finally, we conclude this paper and discuss future work in Section \ref{CON}. 

\section{Experiment Design}
\label{ED}
\begin{figure*}[htbp]
% \vspace{-2mm}
\centering
    \hspace*{-7mm} 
    \includegraphics[width=1.1\textwidth]{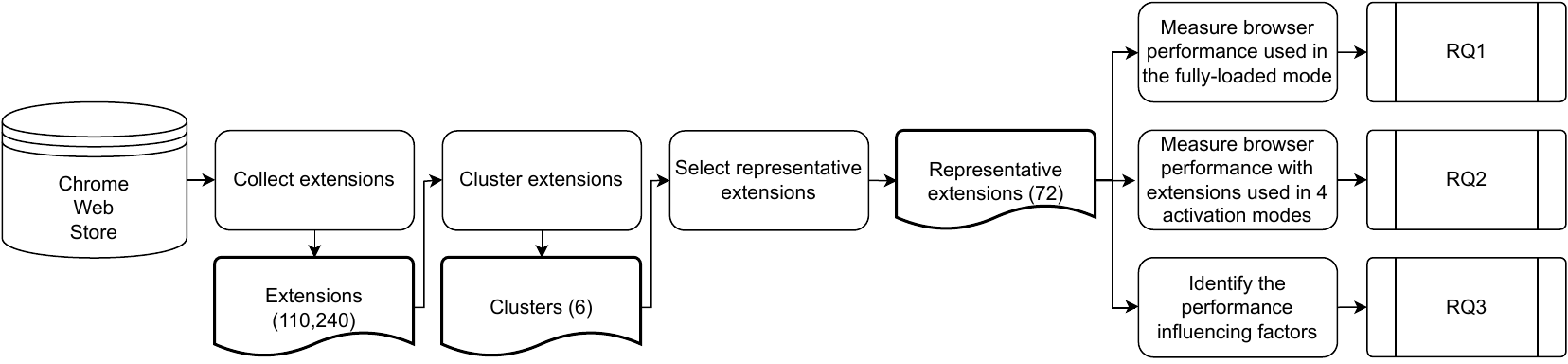}
    % \includegraphics[width=1.1\textwidth]{figures/overview.eps}
    %\vspace{-7mm}

    \caption{An overview of our experiment design and analysis.}
    % \heng{Don't mention ``with privacy practices'' in the second box, also update the box below to include extensions without pp as well. remember you do clustering for all of them (just playing a trick to reduce noise).}}
    \label{fig:flowchat}
    %\vspace{-4mm}
\end{figure*}

% To investigate our defined research questions, we analyze the energy-related posts on Stack Overflow.
%In this section, we outline our data collection, extension selection, testing scenario design, testbed, and measurement and evaluation criterion. 
Figure \ref{fig:flowchat} presents an overview of our approach.
We first collect extensions from the Chrome Web Store, then we cluster the extensions according to their similarities and select representative extensions from each cluster as our final studied extensions.
With the selected extensions, we design experiments to understand how they impact the browser performance
in terms of energy consumption and page load time (RQ1), how different activation modes impact the browser performance (RQ2), and what factors of the extensions impact the browser performance (RQ3).
% In the rest of the sections, we discuss the details of our experiment design, including our data collection, extension selection, testing scenario design, testing procedures, and measurement and evaluation criterion.

% \heng{The experimentation design section needs better organization, following a logical order. Better have an overview figure.}

% \subsection{Overview}
% \heng{add an overview subsection to discuss the overall experiment design.}

% Figure \ref{fig:flowchat} presents an overview of our approach.
% methodology employed in our quantitative study. Our approach comprises of three steps: 1) case study, 2) evaluation, and 3) identification of performance-affecting factors. 
% In particular, the first step
% We first categorize the extensions into six types and identify six activation modes applied to the extensions.
% The selected extensions are evaluated in each activation mode of extensions, with statistical analysis performed on the change ratio of performance. 
% \bihui{goal of your statistical
% analysis}
% Lastly, we investigate the factors that contribute to the impact of the extensions on browser performance.
% The case study provides us with a more detailed understanding of the effect of extensions' activation with the extensions in real-world scenarios.
% As discussed in Section \ref{ER}, the findings are derived from both the case study and the reports on performance-affecting factors.

\subsection{Collecting Extensions}\label{CE} 
\begin{figure}
\centering
    \includegraphics[width=.9\textwidth]{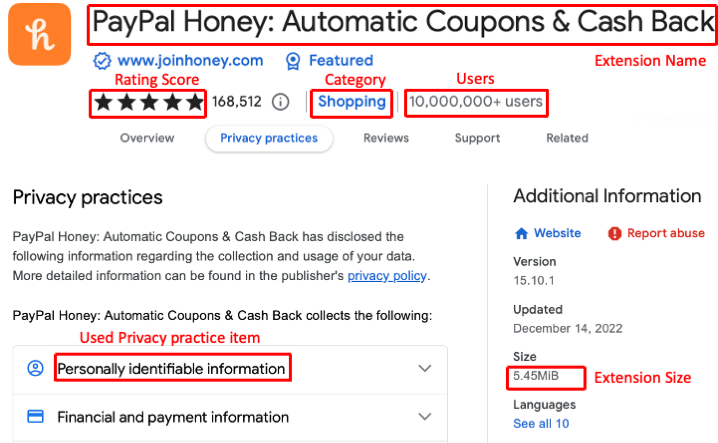}
    \caption{An annotated screenshot of an extension from the Chrome Web Store.}
    \label{fig:ext-overview}
    % \vspace{-4mm}
\end{figure}

Unlike the desktop-based Chrome, the mobile Chrome does not support extensions because mobile devices have limited processing power and storage capacity to support resource-demanding extensions. % compared to desktops. 
Therefore, our study targets the desktop-based Chrome, which can be used in portable devices, such as laptops.
We collect a total of 110,240 extensions across 11 different categories from the Chrome Web Store. 
% To consider the impact of the privacy practices on Chrome's performance (in RQ3), we exclude the extensions without providing privacy practices where the property for privacy practices is marked as "not provided" or "none".
% In the end, 12,423 extensions are gathered.
The collected information of the extensions includes the extension name, category, rating score, used privacy practices, extension size, the number of raters, and the number of downloads.
An example screenshot of an extension is shown in Figure \ref{fig:ext-overview}.
The descriptive information (e.g., star ratings, category, and number of users) is annotated in the example.
% We collect the information of the 12,423 extensions including extension name, category, rating, privacy practices, size, and the number of downloads.

\bihui{Privacy practices of extensions specify the actions and policies that extensions take to collect the personal information and user data. 
Extensions can be useful to enhance the functionality of the Chrome browser but can also pose a threat to user privacy and data security. 
It is important for users to be aware of the privacy implications of the extensions they install. Thus, Chrome Web Store suggests developers to be transparent about their data collection and use practices by declaring their privacy practices in the Chrome Web Store}.
Developers document privacy practices to inform users how they collect, use, store, and share their data. 
Nine types of collected data, including Location, User Activity, Website Content, Personally Identifiable Information, Authentication Information, Web History, Personal Communications, Financial and Payment Information, and Health Information, are documented during the development of extensions and are used to describe privacy practice properties.
\bihui{For example, the use of location services in extensions can track the user's GPS location and is useful for certain types of applications, such as navigation or weather forecasts. 
User Activity monitors the user's activity when users browse the webpage for the purposes of analytics and personalization.}
% A screenshot of an example extension is shown in Figure \ref{fig:ext-overview}. 
% The different components of an extension are annotated in the figure\heng{this sentence is not needed anymore?}.

\vspace{-1em}\subsection{Selecting Representative Extensions}
\bihui{The performance behavior of popular extensions may not be generalized to other ones. 
Thus, we want to extend the existing work (e.g., \cite{Pearce, Merzdovnik, Borgolte}), which studies a few selected extensions, with a more diverse set of extensions instead of considering the popularity of extensions solely.}
% \bihui{R1.14 - Users might naturally lean towards popular extensions when making their decisions, assuming that a larger user base could indicate greater performance, widespread acceptance, and potentially more robust features.}
% \heng{Not sure what the purpose of mentioning popular extensions here is, as this paragraph never touches on popularity. Do you want to justify why you did not consider the most popular extensions? If so, we need to give a rationale, such as the performance behavior of popular extensions may not generalized to other ones. Thus we want to consider a more diverse set of extensions ... \bihui{-solved}}
% Popularity could be an intuitive factor in choosing extensions, as popular extensions may have a larger user base and more contributors. %Therefore, popular extensions can result in more frequent updates and bug fixes, making the extensions more reliable, stable, and energy-efficient.
In particular, we are interested in investigating all types of extensions in order to generalize the performance impact of extensions on the browser.
% \heng{this contradicts with one earlier sentence which says we look at popular extensions -\bihui{changed to only abt all types of extensions}}. 
We use two criteria to select the extensions to ensure their representativeness: (1) we consider the use of privacy practices in extensions to select the studied extensions, as privacy practices (e.g., Location) may have a significant impact on extension performance; (2) \bihui{we include the extensions from different categories}. 
% and with different settings of types of collected data. 
\bj{Due to the enormous of extensions available, it is impossible to study all extensions. 
In our work, we cover a wide spectrum of extensions in all 11 categories of extensions.}
We cluster the extensions with respect to their privacy practices, and then we select one extension from each category for each resulting cluster.
%\bj{For 12,423 extensions that uses privacy practices\heng{comes from nowhere. Check suggested writing above.}, we decide to choose representatives on the basis of their uses of privacy practices, as we want to consider the impact of the privacy practices on Chrome’s performance (in RQ3).}

\textbf{Clustering extensions based on privacy practices.}
\bihui{The majority (i.e., 60\%) of the extensions do not disclose the use of privacy practices}. 
\bj{We exclude such extensions and only consider the extensions with privacy practice specifications to perform the clustering, because the extensions without clarifying privacy practices are untraceable in terms of structure for the analysis.} 
%To eliminate the bias towards popular extensions and ensures to include lesser-known extensions, 
Specifically, we apply the K-medoids clustering algorithm \citep{kmedoids1,kmedoids2} to cluster the 12,423 extensions with privacy practice specifications.
We treat the extensions without privacy practices adopted (i.e., marked as ``none'' in privacy practices) as a separate cluster (i.e., cluster 0).
% , as shown in Figure \ref{fig:ext-overview}.
% K-medoids clustering algorithm can handle better outliers and noise than k-means clustering.  K-medoids algorithm is less sensitive to the initial starting points, and produce clusters with non-convex shapes. As a result, K-medoids algorithm can provide more accurate clustering of extensions.
The K-medoids clustering algorithm outperforms k-means in handling outliers, noise, and non-convex shapes of clusters \citep{clusterBetter,clusterBetter2}.
%By minimizing the sum of dissimilarities of the extensions in a cluster, the K-medoids algorithm results in more coherent and stable clusters. 
% Thus, the K-medoids algorithm can eliminate the bias towards popular extensions and ensures to include lesser-known extensions that have similar privacy practices \heng{this is the reason of using clustering, not why using K-medoids?}. 
Since properties in the privacy practices are categorical values, we use one-hot encoding~\citep{onehot} to convert the privacy practices of each extension into numerical variables (i.e., 1 means True, and 0 means False). 
%Supposing an extension ($EXT$) is a set with three categorical variables $p1$, $p2$, and $p3$. 
For simplicity, supposing there are three privacy practices: Location ($p1$), Personally Identifiable Information ($p2$), and Authentication Information ($p3$), 
if an extension uses privacy practices $p1$ and $p3$, the corresponding one-hot encoding is (1, 0, 1). 
%\bj{For example, the extension\footnote{\url{https://chrome.google.com/webstore/detail/octotree-github-code-tree/bkhaagjahfmjljalopjnoealnfndnagc?hl=en}} uses Personally Identifiable Information, and Authentication Information when building the functionality. The use of the privacy practices is converted to (0,0,0,1,1,0,0,0,0).}
We use the Elbow method \citep{elbow} with silhouette scores \citep{silhouette} to choose the number of clusters. 
The elbow method chooses the ``elbow" point in the \textit{Silhouette score $\sim$ the number of clusters curve} (i.e., a relatively small number of clusters with a relatively higher Silhouette score).
The optimal K value, i.e., the best number of clusters, is then determined to be 5, with a silhouette score of 0.435.
The silhouette score ranges from -1 to 1. 
The higher value indicates better clustering, and the value of 0.435 suggests a favorable clustering result \citep{Silhouettes}.
% Using both the Silhouette \cite{silhouette} and Elbow methods \cite{elbow}, we select the optimal K value, i.e., the best number of clusters, to be 5 \bj{with a score\heng{what score, based on the Silhouette or Elbow method? it is not clear how you combine these two methods} of 0.435}.
Adding the cluster of extensions without privacy practices adopted, we obtain a total of 6 clusters, as shown in Table \ref{tab:clusters}.

\begin{table}
\centering
\caption{\bihui{The extension clusters and examples. }} 
% \heng{I think we don't need to put this table in the paper. We can just use a smaller table to explain each cluster and give an example: four columns (cluster id, explanation and example, total number of extensions, number of extensions sampled. The longer list could be put in the replication package. \bihui{-solved}}}
\label{tab:clusters}
\begin{threeparttable}
\resizebox{\linewidth}{!}{
\begin{tabular}{c|l|c|c} 
\hline
\multicolumn{1}{l|}{Cluster} & Explanation and Example & \begin{tabular}[c]{@{}l@{}}Total Number of\\Extensions\end{tabular} & \begin{tabular}[c]{@{}l@{}}Number of\\Extensions Sampled\end{tabular} \\ 
\hline
0 & \begin{tabular}[c]{@{}l@{}}None of privacy practice adopted in the extensions\\(e.g., Dark Reader\tnote{1} { })\\\end{tabular} & 31908 & 11 \\
\hline
1 & \begin{tabular}[c]{@{}l@{}} The cluster focuses on the adoption of website content\\in privacy practice items\\(e.g., Wappalyzer\tnote{2} { })\end{tabular} & 4056 & 12 \\
\hline
2 & \begin{tabular}[c]{@{}l@{}}The cluster focuses on the adoption of\\authentication information~in privacy practice items\\(e.g., Octotree\tnote{3} { })\end{tabular} & 1951 & 11 \\
\hline
3 & \begin{tabular}[c]{@{}l@{}}The cluster focuses on the adoption of P.I.I.\\in privacy practice items\\(e.g., Keepa\tnote{4} { })\end{tabular} & 3649 & 13 \\
\hline
4 & \begin{tabular}[c]{@{}l@{}}The cluster focuses on the adoption of \\user activity and website content in privacy practice items\\(e.g., ~TubeBuddy\tnote{5} { })\end{tabular} & 1586 & 11 \\
\hline
5 & \begin{tabular}[c]{@{}l@{}}The cluster focuses on the adoption of\\user activity in privacy practice items\\(e.g., Sourcegraph\tnote{6} { })\end{tabular} & 1181 & 14 \\
\hline
\end{tabular}
}
\begin{tablenotes}
    \scriptsize
    % \item \heng{add numbers to the right places in the table to map the table notes, like what is done in Table 2}
       \item P.I.I. refers to Personally Identifiable Information
       \item The complete list is put in the replication package.\\
    \textsuperscript{1} \url{https://chromewebstore.google.com/detail/eimadpbcbfnmbkopoojfekhnkhdbieeh} \\
    \textsuperscript{2} \url{https://chromewebstore.google.com/detail/gppongmhjkpfnbhagpmjfkannfbllamg} \\
    \textsuperscript{3} \url{https://chromewebstore.google.com/detail/bkhaagjahfmjljalopjnoealnfndnagc}\\
    \textsuperscript{4} \url{https://chromewebstore.google.com/detail/neebplgakaahbhdphmkckjjcegoiijjo}\\
    \textsuperscript{5} \url{https://chromewebstore.google.com/detail/mhkhmbddkmdggbhaaaodilponhnccicb}\\
    \textsuperscript{6} \url{https://chromewebstore.google.com/detail/dgjhfomjieaadpoljlnidmbgkdffpack}
\end{tablenotes}
\end{threeparttable}
\end{table}

\textbf{Selecting representative extensions from each category.}
It is time-consuming to measure the performance of an extension. To include all 11 categories of extensions in the Chrome Web Store, we select one representative extension from each category in each cluster.
 The selected representative extension from each category has the minimum Manhattan distance to the medoid of the cluster.
 \bihui{A medoid is a representative object of a cluster at the center of the cluster whose the sum of Manhattan distance to all the objects in the same cluster is minimal.}
 The distance is calculated by the similarity of uses of privacy practices between an extension and the medoid.
The extensions with the minimum Manhattan distance to the medoid are expected to exhibit the most similar characteristics as the other extensions within the cluster (i.e., being the most representative). %, regardless of the popularity of extensions.}
% \heng{Removed the details about Manhattan distance as it is well known}
% \begin{comment}
% The Manhattan distance is calculated using Equation \eqref{eq:manhattan}:
% \begin{equation}
%          d(EXT_1,EXT_2) = \sum_{i=1}^{9} |p_i - p_i^{'}|\label{eq:manhattan}
% \end{equation}

% \noindent where $|\textbf{p}| = (p_1, ..., p_9)$ and $ |\textbf{p}^{\textbf{'}}| = (p_1^{'}, ..., p_9^{'})$ are two feature vectors of extensions $EXT_1$ and $EXT
% _2$, respectively. $p_i$ corresponds to the use of a privacy practice $i$ in the extension $EXT$. $d(EXT_1,EXT_2)$ denotes the Manhattan distance from $EXT_2$ to $EXT_1$. More specifically, the Manhattan distance is computed using a 9-dimensional vector representing privacy practices. Each dimension corresponds to a specific privacy practice. 
% \bj{In each cluster, we choose one extension that has the minimum Manhattan distance to the medoid of its cluster\heng{distance to what? -\bihui{solved}} in each of the 11 categories. 
% \end{comment}
In total, we select 55 extensions from the 5 clusters of extensions with privacy practice specifications (i.e., clusters 1 to 5, listed in Table \ref{tab:clusters}).
% Consequently, 55 extensions are chosen, i.e., 11 extensions representing each category from a cluster multiplied by 5 clusters.
For the cluster of extensions (i.e., cluster 0, listed in Table \ref{tab:clusters}) that do not adopt any privacy practice (i.e., marked as ``none" in privacy practices), the representative extensions cannot be selected in regard to the use of privacy practices, as all these extensions have the same Manhattan distance to each other in terms of privacy practices.
Instead, we select representative extensions as to their popularity, i.e., we choose the extensions with the highest number of users in each category. 
In case two most popular extensions have the same number of users, we consider the rating scores, followed by the number of raters.
As a result, we select a total of 11 popular extensions without adopting any privacy practice, choosing one from each category. %, thereby ensuring a comprehensive representation.
%In total, we selected 66 extensions at this stage.

\begin{figure}
\centering
    \vspace{-6mm}
    \hspace*{-1cm}
    \includegraphics[width=0.7\textwidth]{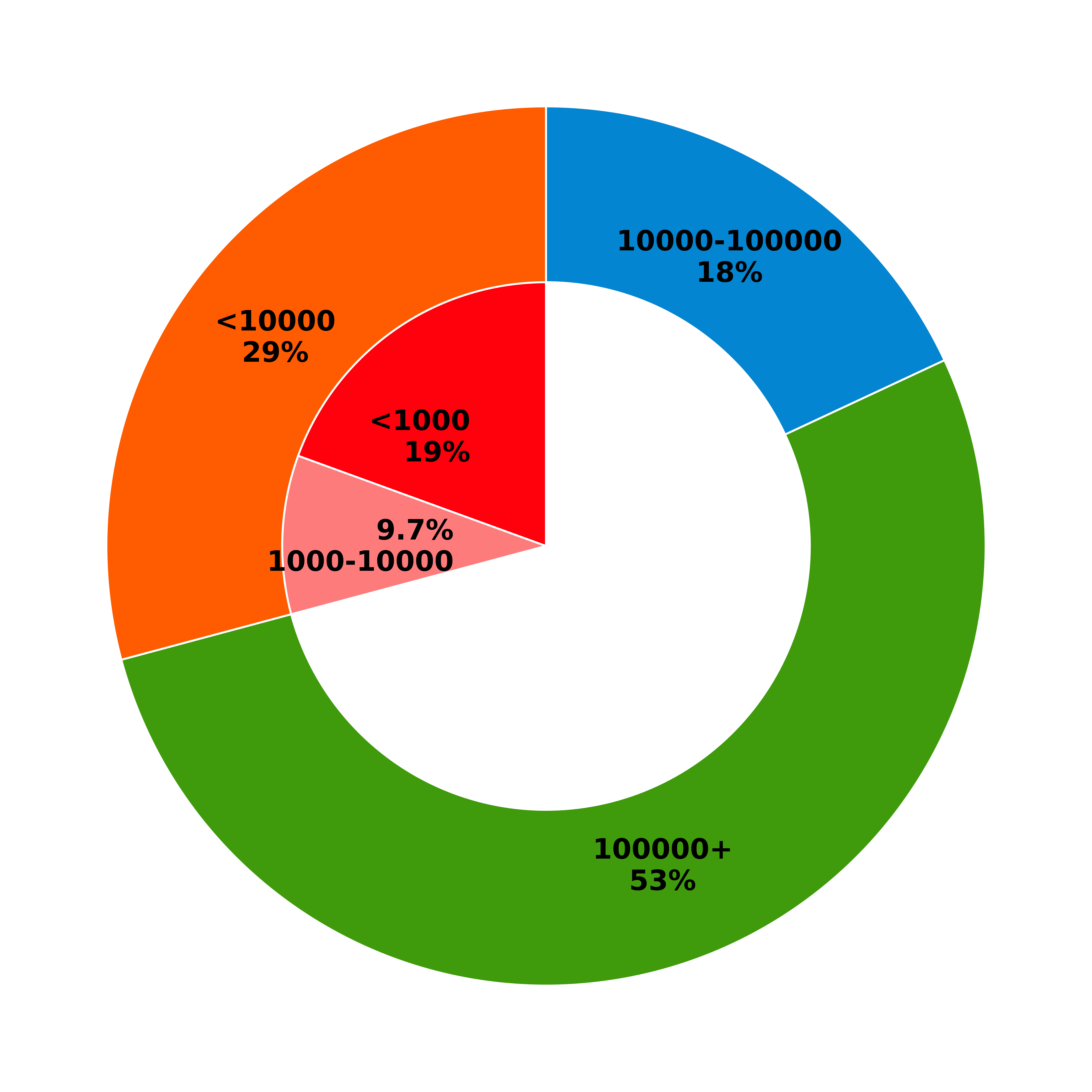}
    \vspace{-6mm}
    \caption{Extension Popularity.}
    \vspace{-4mm}
    \label{fig:ext-pop}
\end{figure}

%Not every extension is suitable for performance measurement.
% However, the extensions possess unique functions and characteristics. 
%For instance, MyJDownloader\footnote{\url{https://chrome.google.com/webstore/detail/myjdownloader-browser-ext/fbcohnmimjicjdomonkcbcpbpnhggkip?hl=en}} is evoked without using web resources in the web browser. 
%As we want to measure the changes in the browser performance rather than the changes in the resources outside of browser, we exclude such extension types.
%We manually evaluate each representative extension and filter out the extensions that are not suitable for measurement. 
% We find 6 extensions that do not provide access to their functions using web browsers.
There are 6 extensions that are not designated for browsing a website, such as Weather\footnote{\url{https://chrome.google.com/webstore/detail/iolcbmjhmpdheggkocibajddahbeiglb}}. 
%To ensure that our study covers all types of extensions, 
To improve the representatives of our selected extensions,
% we complement additional six extensions.}
% Nonetheless, they still use resources on the browser, such as Weather\footnote{\url{https://chrome.google.com/webstore/detail/weather/iolcbmjhmpdheggkocibajddahbeiglb?hl=de}}. 
we retain these 6 extensions and complement them with 6 additional extensions (i.e., ones that work with web browsing) that are closest to the medoid within the same cluster.
% into eight scenarios, as outlined in Table \ref{tab:cluster exception}.
% These scenarios are described in detail, including descriptions, examples, reasons, and solutions, based on the working conditions, functions, and characteristics of the extensions. 
% To facilitate the measurement of the extensions, the extensions that fall into six of the eight identified scenarios, including Build-in Webpage, Foreign Language, External App Support Required, No Actual Function, Subscription Required, and New Tab, are not taken into consideration. Moreover, six extensions that fall under three of the eight identified scenarios are retained and then substituted by the consegment extension that is closest in proximity to the medoid within the same cluster.
As a result, a total of 72 representative extensions are selected,  comprising 11 extensions that do not have privacy practice specifications,
% \heng{Be precise: are they extensions that do not employ privacy practices, or do not have privacy permission specifications? Be precise and be consistent throughout the paper}
55 extensions that implement privacy practices, and an additional 6 complementary extensions. \bihui{Figure \ref{fig:ext-pop} shows the popularity of the chosen 72 extensions, with 71\% of extensions (i.e., 51) having been downloaded over 10,000 times.}
\bj{Table \ref{tab:downloads}, listed in Appendix \ref{Appendix1}, details the number of downloads for each selected extension.}
% in which the oldest version of the extension\footnote{\url{}} (??? downloads) can be traced back to DD/MM/YYYY (DD days from the experiment conducted).}

\subsection{Designing Testing Scenarios}\label{test case design}
% The purpose of the test cases is to evaluate and compare the performance of different extensions in a standardized and controlled environment.
% The test cases are tailored to suit the characteristics of the extension, such as operating conditions, service purpose, and design purpose.
% Through in-depth analyses, six types of extensions are identified, namely {\tt Generic}, {\tt Video Based}, {\tt Shopping Based}, {\tt GitHub Based}, {\tt News Based}, and {\tt Targeted} types, in Table \ref{tab:scenario} to facilitate measurement of the extensions and comparison of their performance changes.
% The test cases ensure that the experiment are consistent and that the results of the performance evaluation are representative of the actual performance of the extensions in real-world conditions.

To measure the performance impact of extensions, we group the extensions based on the similar execution conditions on a website. 
\bihui{For instance, extensions Octotree and Sourcegraph 
% \heng{provide their names: the extensions X and Y... -\bihui{solved}}
share similar working conditions and are intended for use on GitHub, thus we group them together.
Afterward, we establish a testing scenario tailored to such extensions.}
In total, we create seven types of testing scenarios, as listed in Table \ref{tab:scenario}. 
For instance, 40 extensions are universally compatible \bj{(i.e., can be used for all websites)} with all types of websites on the internet. 
% we select the top 10 most popular websites from Semrush (a platform provided the traffic analytics tool) and 
The testing scenario for these extensions is named as the {\tt Generic} scenario.
% For extensions working on video sites, we choose 6 extensions for YouTube and 2 extensions for the Twitch website. 20 videos (i.e., 10 from YouTube and 10 from Twitch) are chosen. To keep consistence among videos, we choose the videos with 720P quality and videos with 10 minutes from YouTube. We also choose 10 videos with 720p quality and 8.5-hour from Twitch. Twitch offers live streaming services. The videos are used for recording. Therefore usually have longer duration. 
% \bj{merge to table 1 -
% The {\tt Video} type tests extensions for video sites, while the {\tt Shopping} type is designed to test extensions that enhance the shopping experience. }
% We classify testing scenario that is designed to test extensions on the video site as the {\tt Video} type.
% The extensions designed for video sites, such as YouTube, are classified as the {\tt Video Based} type.
% The left 10 extensions are categorized into the {\tt Shopping Based} (6), {\tt GitHub Based} (2), and {\tt News Based} (2) types, respectively. 
% Extensions designed to enhance the user shopping experience are classified as  the {\tt Shopping} type. 
% For the extensions designed to enhance the user shopping experience, we select a diverse range of websites, including the 10 best-selling products from each of Amazon's top 10 categories, and classify them as the {\tt Shopping Based} type. 
% The shopping websites used are a selection of the 10 best-selling products from each of Amazon’s top 10 categories to ensure website diversity.
The {\tt GitHub} scenario tests extensions that are used to improve the user experience on the GitHub website.
% We classify testing scenario that is used to test extensions that enhance the user experience on the GitHub website as the {\tt GitHub} type.
% 2 extensions are designed to enhance the user experience on the GitHub website.
We select only code development projects based on the star rankings available on GitHub. 
% GitHub projects are selected for the extensions that enhance the user experience on the GitHub website based on repository ranking. 
% These projects are chosen from code development projects rather than plain-text projects, such as 996icu\footnote{\url{https://github.com/996icu/996.ICU}}, that do not contain code elements.
% We classify testing scenario that is designed to test extensions for news resource websites as the {\tt News} type. % we select five pages from each press website Naver and Daum and classify them as the {\tt News Based} type.
% Since the extensions designed for news resource websites predominately function on the websites Naver and Daum, we select 5 pages from each press website.
For eight extensions that cannot be grouped, 
% \heng{Maybe better call them ``others'', as the ones ``GitHub'', ```Video'' etc. also only work for targeted websites -\bihui{Do u mean ``others'' extensions instead of eight extensions?}}
we classify their testing scenarios as the {\tt Others} type, comprising of eight individual testing scenarios for the eight extensions.
The testing scenarios are summarized in Table~\ref{tab:scenario}.
For each extension in a scenario, we perform the testing with 10 different websites. \bj{For example, we choose the 10 most popular websites from \citet{Semrush} as web content for testing the {\tt Generic} scenario and select the 10 best-selling products in each of \citet{Amazon}’s top 10 categories for the {\tt Shopping} scenario.}

\begin{table}
% \vspace{-2mm}

\centering
\caption{Classification of testing scenarios and the corresponding number of extensions.}
% \vspace{-3mm}
\label{tab:scenario}
% \resizebox{\linewidth}{!}{
\begin{threeparttable}
\begin{tabular}{l|c|l} 
\hline
\multicolumn{1}{c|}{Scenario} & \multicolumn{1}{c|}{\# Ext.} & \multicolumn{1}{c}{Web content used for testing}                                                                                                                        \\ 
\hline
Generic                       & 40                           & 10 most visited websites selected from Semrush\tnote{1}~                                                                                                    \\ 
\hline
Video                         & 10                            & \begin{tabular}[c]{@{}l@{}}10 videos that have 720P resolution and last approximately 10\\minutes from YouTube or approximately 8.5 hours from Twitch\\(depending on the designated websites) \end{tabular}   \\ 
\hline
Shopping                      & 7                            & \begin{tabular}[c]{@{}l@{}}Web pages that host 10 best-selling products selected from each\\of Amazon's top 10 categories\end{tabular}                                  \\ 
\hline
Sport                      & 3                            & \begin{tabular}[c]{@{}l@{}}10 random player profile pages from ESPN\tnote{2}\end{tabular}                                  \\ 
\hline
GitHub                        & 2                            & \begin{tabular}[c]{@{}l@{}}10 projects that contain around 20 lines of code selected\\based on the star ranking from GitHub~\end{tabular}                                  \\ 
\hline
News                          & 2                            & 10 random press articles: 5 from Naver\tnote{3} { } and 5 from Daum\tnote{4}                                                                                                   \\ 
\hline
Others                      & 8                            & 80 designated websites: 10 designated websites of each extension                                                                                                  \\
\hline
\end{tabular}
\begin{tablenotes}
    \scriptsize
    \item \textsuperscript{1} \url{https://www.semrush.com/blog/most-visited-websites};  
    \textsuperscript{2} \url{https://www.espn.com}; \\
    \textsuperscript{3} \url{https://news.naver.com};
    \textsuperscript{4} \url{https://news.daum.net}.
\end{tablenotes}
\end{threeparttable}
% }
\vspace{-2mm}

\end{table}

\vspace{-4mm}
\subsection{Performance Measurement}
% \heng{updated the motivation for the studied properties}
% When a browser takes longer to load a webpage, it could be frustrating and impair productivity \citep{Tian, Borgolte}. Energy consumption is another concern for browser users, particularly for users of battery-powered devices (e.g., laptops)~\citep{Banerjee, Kor}, while the energy consumption of electricity-power devices (e.g., desktops) could also lead to significant environmental impact \citep{Amsel, San} with the excessive power consumption.
% The current studies on energy consumption tend to predominately focus on CPU usage while neglecting the actual patterns of energy consumption resulting from software \citep{Tiwari, Amsel, Fei}.

We study the impact of extensions on the page load time and energy consumption of browsers. 
% We measure the performance of a browser using the following three metrics: 
% \heng{Merge Measured Attributes here and change all ``measured attributes'' to ``performance metrics''}
% As shown in Figure \ref{fig:ext-explain}, 
More specifically, \textbf{the page load time} is the time duration that takes for the webpage to load completely and is measured in seconds. 
% The response time of payload is an important metric to measure the performance of a website, as 
A long page load time (i.e., slow response time) can result in a poor user experience.

% \heng{moved above, not needed here anymore}\bihui{Energy consumption is a concern for many devices, including both desktops and laptops where the studied Chrome version is compatible with studies \cite{Macedo,Pearce,Borgolte}.}\heng{some references -\bihui{I found some studies  used Chrome to measure in desktops or drawn a conclusion that energy consumption is a concern for Chrome.}} 

We measure two stages of energy consumption as follows:

\begin{itemize}[leftmargin=*,topsep=0pt]
\item\textbf{The page load energy consumption} measures the amount of energy in joules (a unit of energy) consumed by the CPU and RAM  by the entire system during the page loading time. 
%As energy consumption is a concern for many devices \heng{including both desktops and laptops where the studied Chrome version is compatible with}, it is important to consider page load energy consumption.

\item\textbf{The stabilized energy consumption} measures the energy consumption of the CPU and RAM by the entire system in joules during a fixed period of time after a webpage has been fully loaded. 
\end{itemize}
%The page load performance differs from the stabilized energy consumption by measuring the energy consumption during the loading process. 
% Stabilized performance provides insight into the energy consumption of a webpage once it has been fully loaded, which can help identify any issues with the webpage that may cause excessive energy consumption.
%The stabilized energy consumption  helps identify issues with the webpage that may cause excessive energy consumption.

To collect energy 
% \heng{and page load time?-\bihui{no RAPL is only used to measure energy consumption}}\heng{later you said ``as well as the running time''} 
measurements, we employ Running Average Power Limit (RAPL)~\citep{rapl}. 
% \heng{Make it clear you are measuring system-level performance or only the performance of the browser process.}
%https://web.archive.org/web/20190116164417/https:/01.org/rapl-power-meter
RAPL\footnote{\url{https://www.intel.com/content/www/us/en/developer/articles/technical/software-security-guidance/advisory-guidance/running-average-power-limit-energy-reporting.html}} is well-established~\citep{DGHHL_2010, pereira_influence_2016, MPEC_2015} and leverages hardware performance counters to provide detailed and precise reading on system energy consumption of the CPUs and memory usage~\citep{Giardino, Khan}. 
% RAPL is a well-established and widely used utility that has been employed in related work, such as ~\cite{DGHHL_2010, pereira_influence_2016, MPEC_2015}, and 
Its accuracy has been validated by various studies~\citep{rapl_accuracy_2018, rapl_accuracy_2016, paniego_analysis_2018, KD_2019}. 
%RAPL offers a high sampling interval, with one reading taken per 1 millisecond.
The results obtained from RAPL provide a measurement of the total energy consumption in millijoules. 
% The running time of loading a webpage is recorded in seconds.\heng{From this description it is not clear how you distinguish page load energy and stabilized energy measurement -\bihui{next to this comment}}\bihui{When the webpage finalizes its loading process, RAPL continues recording a measurement of the total energy consumption in millijoules.}

\subsection{Testing Procedure}\label{measurement procedure}

\begin{figure}[h]
\vspace{-2mm}
\centering

    \includegraphics[width=0.8\textwidth]{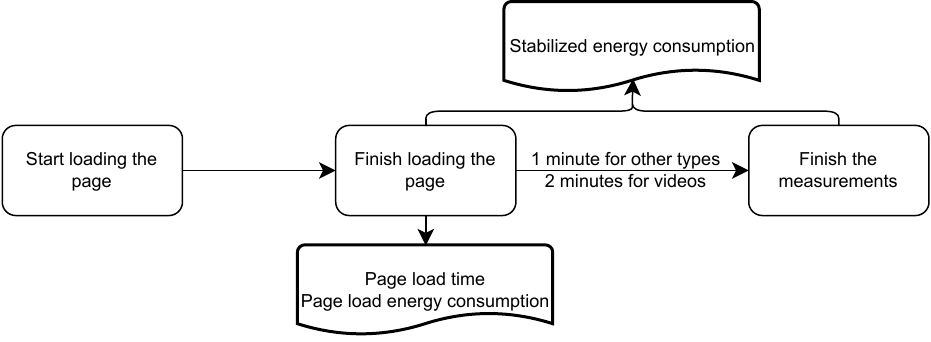}
    \vspace{-1mm}
    \caption{Measurement timeline}
    \label{fig:ext-explain}
    \vspace{-4mm}
\end{figure}

To conduct our experiment, we use a desktop equipped with an Intel i7-4770 @3.4GHz processor, 32 GB of RAM, running Ubuntu (kernel version: 5.15.0-48-generic) with both WiFi and Bluetooth disabled.
% \heng{this goes to Testing Procedure too}
We use Selenium\footnote{\url{https://www.selenium.dev}}, a tool that generates simulation scripts, to automatically execute a web browser to access designated websites. The Google Chrome browser of version 104.0.5112.79 (Official Build 64-bit) is utilized in our experiment.

To account for measurement errors, we calculate the minimal number of repetitive measurements for each performance metric to achieve an acceptable level of accuracy, using Equation~\ref{eq:sample-size}~\citep{sample-size}.

\begin{equation}
         \textnormal{n} = \lceil{(\frac{100 \times z_{1-\alpha/2} \times s}{r \times \bar{x}})}^2\rceil
         \label{eq:sample-size} 
\end{equation}

\noindent where $n$ stands for the number of observations required; $z_{1-\alpha/2}$ is the $1-\alpha/2$ critical value of the normal variate at the desired confidence level ($1-\alpha$);
% , where $\alpha$ is the probability of rejecting a null hypothesis when it is true
$s$ represents the median standard deviation of the measurements for each website and extension combination; $r$ denotes the required accuracy; and $\bar{x}$ represents the median value of the sample mean of the measurements for each website and extension combination.
% \heng{or for each website and extension combination?\bihui{-solved}}.

\bihui{With a 95\% confidence level and a required accuracy of 10\%, we obtain the minimal number of repetitive measurements for the performance metrics based on Equation \eqref{eq:sample-size}: 9 times for the page load time, 6 times for the page load energy, and 1 time for the stabilized energy consumption. 
The different numbers of repetitive measurements stem from the inherent characteristics of each metric and the impact of measurement errors. 
For instance, page load time might exhibit more variability, necessitating a larger number of measurements to achieve the desired accuracy level. 
Stabilized energy consumption, being more stable, requires fewer measurements to meet the accuracy criteria. 
In our experiment, we choose an ample number of repetitive measurements (i.e., 10) that is larger than the largest number required for any of the performance metrics (i.e., 9).} 

We use scripts to open and navigate to the designated web pages and simulate users' activities in practice.
To measure each testing scenario, we perform the experiment in the following steps: 

% \begin{itemize}[leftmargin=*,topsep=0pt]

  \textbf{1)} Utilize Selenium to launch the Google Chrome browser.

 % \item \heng{Missing a step to install/activate \& configure an extension?}
  \textbf{2)} Install and configure an extension on demand.

  \textbf{3)} Repeat each test case (i.e., each website under each scenario) 10 times to obtain stable measurements in that testing scenario, as specific energy measurements may vary across executions of the same testing scenario. For each run of a testing scenario, the local storage and caches of the browser are cleaned to avoid that the caches would maintain the results of the previous experiments. In total, we test and measure the performance of each extension 100 times (i.e., 10 websites, each with 10 repetitions \citep{10times}).%We execute the testing scenario for each extension 10 times (i.e., 100 measurements in total for 10 websites). 
 
  \textbf{4)} Collect performance measurements of energy and running time while the testing scenario is running, as illustrated in Figure \ref{fig:ext-explain}.
  % \heng{missing figure? \bihui{-moved up}}.
 
  \textbf{5)} Sleep for a minute to avoid tail power states~\citep{Bornholt} to allow the system to reach a stable condition again (idle energy consumption) before executing the next run (i.e., running the browser to access one website).

  \textbf{6)} Terminate the Google Chrome browser for each finished task by closing all windows and pages, thereby ending the browser session that is opened during measurement.

  \textbf{7)} Uninstall the current extension.
  
   \textbf{8)} 
 % \heng{Uninstall/deactivate the current extension first?}
 Repeat 1) to 7) for another extension.

 \textbf{9)} Collect the energy consumption of the CPU and memory, and the page load time on each website without any extension.
% \end{itemize}

% For each finished measurement, we reset the browser by cleaning the local storage and cashes. 
% For each finished task, we reset the browser by closing all windows and pages, i.e., terminating the browser that is opened during measurement.
The page load time and the page load energy consumption are monitored from the time the webpage starts loading  until the webpage completes the loading. 
%Once the page finishes loading, the energy consumption of CPU and memory usage is measured for two minutes in the {\tt video} type of testing scenarios and one minute for other testing scenario types. 
Once the page finishes loading, the stabilized energy consumption of CPU and memory usage is measured for one minute for non-video testing scenarios and two minutes for the {\tt video} testing scenario. % and one minute for other testing scenario types. 
We choose the one-minute testing time to reach a balance between the duration of the measurement and the time resources available for executing all the testing scenarios (it takes approximately 10 hours to execute the testing scenarios).
The two-minute measurement duration for the {\tt video} scenario is intended to account for the ad time in videos.
% and measure only
% \heng{is it true that you removed the energy consumption caused by ads? -\bihui{I'm sorry for the misunderstanding. The measurement includes the energy consumption caused by ads.}} 
% the energy usage of the video content.
%To obtain stable measurements in other testing scenario types, we monitor the energy usage of the website for one minute.

% \heng{Moved the explanation about the sample size here \bihui{-thank you}}: 

\bihui{By repeating each experiment 10 times, we obtain stable measurement results across the page load time, page load energy consumption, and stabilized energy consumption. These measurement results exhibit good stability, as evidenced by median standard errors \citep{ste} of 0.050, 0.042, and 0.0035 (calculated from measured metric values normalized by median normalization in  Section \ref{normalization}) for the respective metrics. 
In the experiment, Selenium initially opens a blank webpage to avoid caches and the impact from previous runs on the measurements, before the initial page is directed to the testing webpage.}

\subsection{Normalization} \label{normalization}
% Based on the measurement collected in the extension-free mode for each of test cases, the values in each test case are normalized with respect to each site in the test case. Normalization for the measurements of the extension-free mode is followed by Equation \ref{eq:ext-normalization}. Normalization is computed by the type of the test case and based on the website.
% To ensures that all measurements collected during the testing of extensions, regardless of the mode or test case, are consistent and comparable, we normalize the measurements obtained with and without the extension used. 
To ensure consistency and comparability of all measurements collected during the testing of extensions, regardless of the testing scenarios and the websites, we normalize the measurements obtained with and without the use of an extension.
%The normalization process is guided by Equation \eqref{eq:ext-normalization} and is calculated based on the testing scenarios
% \heng{not type of testing scenarios, just testing scenario is fine} 
%and the websites.
For each extension, the normalized value of the $j$th ($j$ $\in {1, 2,.., 10}$) measurement of the browser's performance when accessing the $i$th ($i$ $\in {1, 2,.., 10}$) website ($\textnormal{perf\_norm}_{ij}$) is calculated by Equation \eqref{eq:ext-normalization}. 

\begin{equation}
         \textnormal{perf\_norm}_{ij} = \frac{\textnormal{perf\_ext}_{ij}}{\textnormal{median}(\textnormal{perf\_free}_i)}
         \label{eq:ext-normalization} 
\end{equation}
% \heng{seems wrong here: you normalized for each website, nothing to do with type right? then why you have type here?} \heng{should highlight that it is normalized by the extension-free mode. Current equataion is confusing.}
\noindent where $\textnormal{perf\_ext}_{ij}$ is the raw value obtained by the $j$th measurement of the browser performance while the extension accesses the $i$th website; and $\textnormal{median}(\textnormal{perf\_free}_i)$ is the median value obtained by the 10 measurements of the browser performance (free of extension) when the extension accesses the $i$th website.
Similarly, the performance measures of the browser in the extension-free mode are normalized by the median value of the 10 measurements for each website. 

\section{Experiment Results} 
\label{ER}
The objective of this study is to investigate the impact of extensions on the energy consumption and page load time of browsers through a case study on Google Chrome.
This section describes our research questions, our approaches, and answers to each of them.

%\textbf{a) 
\subsection{RQ1: How do extensions impact browser performance?}\label{rq1}

\textbf{\underline{Motivation:}}
% Google Chrome, as the most widely used modern browser connecting users to the Internet, enhances the user surfing experience by supporting the installation of extensions. 
Extensions are developed to enrich the browser functionality, such as easy dictionary searching or blocking advertisements. %, or catering to the demands of users. 
However, practitioners may not be aware that extensions may cause significant performance impact on browsers, %consume energy from the computer in various ways, 
such as extra CPU computation and memory usage, and increased the page loading time. 
Adversely, extensions could result in degraded user experience and impaired performance of other applications running on the same computer. 
% Using extensions can have a significant impact on energy conservation and ecosystems\heng{avoid saying such things as it is a question for this RQ to find the answer}.  
In this RQ, 
% we investigate the effect of extensions on the overall browser performance.
we strive for understanding the extent to which extensions affect the overall browser performance.
% and provide insights into the adoption of energy-efficient extensions\heng{do we provide such insights in this RQ?}.

\noindent\textbf{\underline{Approach:}}
% \heng{this sentence is about details, can be put after describing the main approach}
We conduct performance evaluations of 72 representative extensions by executing their corresponding testing scenarios (Section \ref{test case design}). 
For each extension, we execute the designated websites 10 times with and without the extension installed
% \heng{be precise what ``without ... enabled'' or ``disabled'' mean: deactivated or uninstalled?} 
(Section \ref{measurement procedure}). Other extensions are uninstalled.
% \heng{be clear what this means}.
Measurements are then collected and normalized, as described in Section \ref{normalization}.
Subsequently, we perform statistical analysis to understand the performance impact of the extensions.

%\heng{It reads weird to have this paragraph (before ``In this RQ...'') in RQ1. Move it to Section 2 – Experiment Design, as a separate sub-section after 2.3 - Designing Testing Scenarios.}
%\noindent\textbf{Configuring activation modes of extensions.} Extensions may have different performance impacts depending on the activation modes.
% \heng{Simplified this paragraph to not mention other activation modes.}
\noindent\textbf{Configuring the extensions.} Different configurations of an extension may have different performance impacts\footnote{The performance impact of the extensions' different configurations (i.e., activation modes) is discussed in detail in RQ2.}.
%We consider six activation modes of extensions, listed in Table \ref{tab:user mode}\footnote{The details of the activation modes are discussed in RQ2 where we discuss the impact of the activation modes of extensions on browser performance.}.
%\bj{The activation modes of extensions can be classified based on the conditions for accessing and executing an extension, including whether login to an extension is required, whether permission to access the website is granted, whether an extension is required\heng{required or able?} to run on the targeted website and the usage\heng{not clear what this means} of an extension on the browser.}
In this RQ, we consider the expected situation %primarily focus on the performance changes 
when an extension is used in the fully-loaded mode: the extension is activated and used for the designated websites;
% This is because this RQ aims at studying overall performance changes, and 
%which fully utilize the support of the web browser. 
for extensions
% \heng{maybe only keep ``extensions'' here to avoid more confusion} 
that require access and login, we %perform the ``agree'' operation, which 
grant the extension access and logs the extension in.
% \heng{The two following sentences do not belong to this paragraph. Maybe move and merge with the first paragraph? }
% \heng{not clear, is the extension-free mode different from the baseline?}.
% \heng{Also mention the baseline (extension not enabled). Make it clear you are comparing two scenarios (with and without extensions).}
As the baseline, we also measure the performance of the browser when the extensions are uninstalled (i.e., extension-free mode).
% \heng{disabled or not installed? Also it needs to be clear in Table 3: what does it mean by "No extensions are used"? Are they not installed, or installed but deactivated? -\bihui{they are still installed but not active other than the targeted websites (e.g., on the general websites). I've also added a column to Table 3 called extension deactivated}} (i.e., \textit{extension-free} mode).}\heng{then don't call it ``uninstalled'', maybe ``disabled?''? Also fix other places like the steps in 2.5} 

\noindent \textbf{Statistical analysis.}
To quantify how extensions affect the browser performance, we test the following hypothesis:

% \noindent \textbf{1) Pairwise comparison of the distribution of metric values.}
% \bihui{To study the distribution of performance energy metric values in the two activation modes (i.e., fully loaded mode and extension-free mode), we analyze the metric values of all extensions in each performance energy metric. 
% To examine the overall impact of extensions $e$ on a performance energy metric $m$, we test the following null hypothesis for every pair of groups (e.g., the page load time under the extension free and the fully loaded mode, respectively) divided by extensions $e$.}

$H0_1$: \textit{there is no difference in the distributions of performance metric values between the paired observations of extension-free and fully-loaded modes.}

To compare the two distributions of performance metric values (i.e., fully-loaded mode vs. extension-free mode) for each extension, %$m$ values between two groups of extensions operating in the fully loaded mode and the extension-free mode,

We apply the Wilcoxon signed-rank test \citep{wilcoxon} at a 5\% significance level to assess whether a statistically significant difference exists. %there is a significant disparity in the paired performance energy metrics' distributions. 
The Wilcoxon signed-rank test, as a non-parametric statistical test, does not assume a normal distribution. %If we observe a statistically significant difference (i.e., p-value $<$ 0.05), we reject the null hypothesis and assert that \bj{there is statistically significant difference in the distributions of performance energy metric values between extension-free and fully loaded modes}.
If $H0_1$ is rejected (i.e., a statistically significant difference exists), we further compute the Cliff's $\delta$ effect size (a non-parametric method without assumption of a particular distribution)~\citep{Cliff}, which
% \noindent \textbf{2) Quantifying the importance of the difference.}
%The Cliff's $\delta$ test 
quantifies the magnitude of the differences~\citep{effect-size}. 
%\bihui{As Cliff’s $\delta$ estimates non-parametric effect sizes, it makes no assumptions of a particular distribution.}
% Effect size is used to determine the practical significance of the outcome of the results. The effect size measures the difference between pairs, and is a way to determine if a research finding has practical relevance. 
%A large effect size indicates that the results have practical significance, while a small effect size suggests that the results have limited practical applications.
% We only perform the Cliff's $\delta$ test for a performance energy metric of an extension when the Wilcoxon signed-rank test indicates that an extension leads to a statistically significant performance difference between the performance energy metrics of the two tested modes.
%\st{The Wilcoxon signed-rank test is conducted using the python library, called {\tt scipy}. The Cliff’s Delta test is performed using {\tt cliffs-delta} in python.
%The Cliff's Delta test (Cliff's $\delta$) is applied recursively for extensions that exhibit significant differences in performance metrics compared to the baseline measurements, as determined by the Wilcoxon signed-rank test. }
% We report effect size results in our replication package.
% \noindent \textbf{3) Interpreting the effect sizes.} 
% Only the metric values that pass the statistical tests are deemed to possess a statistically significant difference and hold practical meaning.
The resulting effect sizes are classified into several qualitative degrees of difference \citep{cliff_interval}: negligible ($|\delta|<0.147$), small ($0.147\leq |\delta|<0.33$), medium ($0.33\leq |\delta|<0.474$), and large ($|\delta|\geq 0.474$). A larger effect size signifies a larger difference between the two distributions.
%\bihui{A small effect size implies limited practical applications, while a large effect signifies that the results hold practical significance and that the differences are readily noticeable to a careful observer.}

\noindent\textbf{Performance change ratio.}
%The change ratio for the measurements that show statistically significant differences in the Wilcoxon signed-rank text and practical significance in the Cliff’s Delta test compared to the baseline (i.e., extension-free mode) is then calculated for each performance metric. % using the procedure outlined in a previous section \ref{change-ratio}. 
% When an extension leads to a statistically significant performance difference between the performance energy metrics of the two tested modes, we calculate the change ratio of the performance energy metric. 
If $H0_1$ is rejected, we calculate the change ratio of the performance metric as specified in Equation \eqref{eq:ext-ratio}. 
% \heng{The following sentences maybe deleted later if space is not enough}The change ratio is determined by comparing the normalized measurements obtained with and without the extension enabled under the corresponding testing scenarios \heng{I feel we don't need to mention ``under the same type of testing scenarios'', as each extension has only one corresponding scenario?}. % for both measurements. 
To reduce the effect of outliers in website measurements, the median value of each 10 repeated experiments is used. 
%\bihui{The change ratio for a performance metric is calculated by taking the median of the 10 extension measurements for the $i$-th website, subtracting the median of the 10 extension-free measurements for the same website, and then dividing by the median of the 10 extension-free measurements for that website. The average of change ratios of all 10 websites is then taken to provide a general assessment of the energy and runtime effect of the extension.}
%The overall difference (siginf.) is determined by taking the average of the change ratios across all effect sizes.
% Specifically, Equation \eqref{eq:ext-ratio} computes the change ratio of an extension in terms of its performance energy metrics: %i.e., the page load time, the page load energy consumption, and the stabilized energy consumption.
\begin{equation}
Ratio(metric) = \frac{1}{10}\sum_{i=1}^{10}\frac{median(ext_i)-median(free_i)}{median(free_i)}
         \label{eq:ext-ratio} 
\end{equation}
\noindent where $Ratio(metric)$ is the change ratio of the corresponding performance metric of an extension % $Ratio$ is the change ratio, $metric$ represents the performance metric 
(i.e., the page load time, the page load energy consumption, or the stabilized energy consumption); $ext_i$ is the normalized performance metric under the fully-loaded mode of the extension; %extension measurements for the related metric, 
$free_i$ is the normalized performance metric  under the extension-free mode of the extension; %measurements of the related metric, 
$median(ext_i)$ and $median(free_i)$ indicate the median of the 10 extension measurements for the $i$-th website, under the fully-loaded mode and the extension-free mode, respectively. Finally, $i$ $\in {1, 2,.., 10}$ indicates the $i$-th tested website for the extension. %, and finally $Ratio(metric)$ is the change ratio of the related metric.

\begin{table}
% \vspace{-2mm}
\Large
\centering
\caption{Statistical performance changes caused by the extensions in their fully-loaded mode.}
% \vspace{-2mm}
\label{tab:fully-loaded}
\resizebox{\linewidth}{!}{
\begin{threeparttable}
\begin{tabular}{l|c|cr|crcrcr} 
\hline
\multicolumn{2}{c|}{\multirow{2}{*}{Statistical analysis}} & \multicolumn{2}{c}{Wilcoxon} & \multicolumn{6}{c}{Cliff’s $\delta$} \\ 
\cline{3-10}
\multicolumn{2}{c|}{} & \multicolumn{2}{c|}{Signif.} & \multicolumn{2}{c}{Small} & \multicolumn{2}{c}{Medium} & \multicolumn{2}{c}{Large} \\ 
\hline
\begin{tabular}[c]{@{}l@{}}\\metrics\end{tabular} & Tendency & Count & \multicolumn{1}{c|}{Ratio} & Count & \multicolumn{1}{c}{Ratio} & Count & \multicolumn{1}{c}{Ratio} & Count & \multicolumn{1}{c}{Ratio} \\ 
\hline
\multirow{3}{*}{\begin{tabular}[c]{@{}l@{}}Page load \\time\end{tabular}} & Increase & 21 & 18\% & 10 & 16\% & 3 & 29\% & 8 & 19\% \\ 
\cline{2-2}
 & Decrease & 16 & -7.2\% & 9 & -4.2\% & 5 & -9.0\% & 2 & -43\% \\ 
\cline{2-2}
 & Overall & 37 & 5.1\% & 19 & 3.5\% & 8 & -4.6\% & 10 & 18\% \\ 
\hline
\multirow{3}{*}{\begin{tabular}[c]{@{}l@{}}Page load\\energy\\consumption\end{tabular}} & Increase & 41 & 17\% & 12 & 9.0\% & 8 & 18\% & 21 & 35\% \\ 
\cline{2-2}
 & Decrease & 4 & -6.6\% & 1 & -1.4\% & 2 & -6.6\% & 1 & -61\% \\ 
\cline{2-2}
 & Overall & 45 & 16\% & 13 & 8.6\% & 10 & 15\% & 22 & 31\% \\ 
\hline
\multirow{3}{*}{\begin{tabular}[c]{@{}l@{}}Stabilized \\ energy \\ consumption\end{tabular}} & Increase & 45 & 2.0\% & 12 & 1.9\% & 10 & 1.3\% & 23 & 3.0\% \\ 
\cline{2-2}
 & Decrease & 6 & -1.6\% & 4 & -0.95\% & 2 & -3.0\% & 0 & - \\ 
\cline{2-2}
 & Overall & 51 & 1.7\% & 16 & 0.61\% & 12 & 0.79\% & 23 & 3.0\% \\ 
\hline
\multicolumn{2}{l|}{\# of extensions being tested} & \multicolumn{8}{c}{72} \\
\hline
\end{tabular}
\begin{tablenotes}
      \large
      \item * This table shows the number of extensions that have a statistically significant effect on each performance metric, including an increase, decrease, and overall change (increase and decrease), as well as the respective change ratio. The Signif. column counts extensions with a p-value less than or equal to 0.05. The change ratios are the mean values of the extensions. %We compare the normalized measurements obtained with and without the extension enabled under the corresponding testing scenarios to determine the mean values.

    %\item * The Signif. column counts extensions with a p-value less than or equal to 0.05. % and at least a small effect size.

     \item * The results of the effect sizes, as determined by Cliff's $\delta$, are categorized as Negligible (not reported in the table), Small, Medium, and Large. 
     The effect sizes are calculated and reported only when the performance impact is statistically significant, i.e., the p-value is less than or equal to 0.05 in the Wilcoxon signed-rank test. % and categorical effect sizes.
     
\end{tablenotes}
\end{threeparttable}
}
% \vspace{-6mm}
\end{table}

\noindent\textbf{\underline{Findings:}}
%\textbf{Overall, the page load time, the page load energy consumption, and the stabilized energy consumption increase when an extension is used, among which the page load energy consumption increases the most.} 
\textbf{Using an extension can either deteriorate or improve browser performance, while performance deterioration (especially large deterioration) is more common.}
Table \ref{tab:fully-loaded} presents the statistically significant changes in the performance metrics when an extension is used in its fully-loaded mode.
% The table reports three performance metrics (e.g., the page load time), showing the number of extensions that statistically increase, decrease, or change (increase or decrease) each performance metric, as well as the corresponding change ratio. 
% The results of the extent of the effect size, as determined by Cliff's $\delta$, are categorized as Small, Medium, and Large, while the Signif. category \bihui{ encompasses all extensions with a p-value less than or equal to 0.05 in the Wilcoxon signed-rank test and includes all three effect sizes together.}
A total of 72 extensions are tested, among which 64 extensions (i.e., 89\%) exhibit a statistically significant performance impact on at least one of the studied performance metrics. %and at least 51.39\% of the extensions (i.e., 37) exhibit a statistically significant performance impact on performance energy metrics. %in the fully loaded mode. 
% We observe that the extensions can either improve or deteriorate the browser performance in terms of each performance energy metric. 
In particular, 37 extensions (i.e., 51\%) exhibit statistically significant changes in the \textbf{page load time}, among which 21 increase the metric value while the other 16 decrease the metric value. %with an average increase of 1.00\%.
45 extensions (i.e., 63\%) exhibit statistically significant changes in the \textbf{page load energy consumption}, among which 41 and 4 lead to an increase and a decrease in the metric value, respectively. % with an average increase of 15.57\%.
51 extensions (i.e., 71\%) exhibit statistically significant changes in the \textbf{stabilized energy consumption}, with 45 and 6 of them increasing and decreasing in the metric value, respectively. %with an average increase of 1.44\%.
% On the other hand, 67 extensions show no statistically significant changes in terms of the response time of payloads (29), energy consumption of payloads (21), and stabilized energy consumption(17), representing 42.62\%, 34.43\%, and 27.87\% of the total 61 extensions, respectively.
\textbf{In particular, performance deterioration dominates the biggest performance changes (i.e., with \textit{large} effect sizes).} For example, 23 extensions lead to \textit{large} increases, but 0 extension decreases in stabilized energy consumption.

The deterioration in browser performance may be attributed to the extra resources required to run the extensions, the automatic backend search capabilities of certain extensions, such as potential coupon detection (e.g., SimplyCodes) and tracking of product price history (e.g., Keepa), web content adjustments (e.g., Octotree), monitoring of website content (e.g., Dynatrace), and site analysis (e.g., Wappalyzer) during page load.
On the other hand, the improvement in the browser performance may be due to the blocking of web content (i.e., Naver/Daum Media Filter and ad blocker: Inforness), simplification of web content (e.g., Better Tab), and preventing information trackers (e.g., Neeva).

% , which reduces the page loading time and the energy consumption of the payload \cite{Pearce}.

%\heng{Explain why? why increase, why drop... Same for the paragraphs below.}

 % \heng{highlight the 18.07\% increase for the ones that ``increase'' instead, like: 16 out of the extensions statistically significantly increase the response time, with an average increase of 18.07\%. Same for the bold sentences below, and the summary box.}
\textbf{21 (29\%) extensions statistically significantly increase the page load time, with an average increase of 18\%, while another 16 (22\%) extensions statistically significantly reduce the page load time by 7.16\% on average.}
Over half (37/72) of the extensions lead to statistically significant changes in the page load time, with an average change of +5.1\% (increase).
%show an overall 1.00\% increase in terms of the page load time compared to the measurements without the extension being used.
%Among the 32 extensions that statistically significantly change the page load time, 16 increase the page load time by an average of 18.07\% while another 16 decrease the page load time by 7.16\% on average. 
In particular, eight extensions lead to an increase in the page load time by a large effect size, while only two extensions lead to a decrease by a large effect size. 
%26.23\% (16/61) of the extensions exhibit an increase in response time of 17.30\% in small, 29.34\% in medium, and 18.70\% in large effect sizes, while 14.75\% (9/61), 8.20\% (5/61), and 3.28\% (2/61) of the extensions demonstrate decreases of 4.16\%, 9.02\%, and 43.39\% in the small, medium, and large difference sizes, respectively. 
%\bj{If a large effect size were observed, we would suggest that (1) extensions exhibiting such a large effect size are statistically significant differences in the performance energy metric; and (2) the difference is important.}
Web browsers or extensions may provide warnings about the extensions' impact on the page load time when users install such extensions, suggest users to disable certain extensions when accessing performance-critical websites, or optimize the extension or browser-extension integration to minimize the negative performance impact.

%\textbf{The use of extensions increases the overall energy consumption by 15.57\% during the payload.} 
\textbf{Using extensions leads to the largest negative impact on the page load energy consumption, compared to other performance metrics.}
% \textbf{41 (56.94\%) extensions  statistically significantly increase the energy consumption during the page load, with an average increase of 16.59\%.}
45 out of the 72 extensions demonstrate statistically significant differences in the page load energy consumption, with an average change of 16\% (increase). 
% \heng{I feel not necessary here as later we highlighted the large effect size}\bj{In particular, 18.06\% of the extensions (i.e., 13) with an increase of 8.59\% in small effect size, 13.89\% (i.e., 10 extensions) with an increase of 14.75\% in medium effect size, and 30.56\% (i.e., 22 extensions) with an increase of 31.29\% in large effect size.}
A statistically significant increase in energy consumption is observed in 57\% of the extensions (i.e. 41), with an extra 17\% of energy on average consumed.
Conversely, only 5.6\% of the extensions (i.e., 4) reduce energy consumption, by 6.6\% on average. 
In particular, 21 (29\%) extensions exhibit a large effect size in the increase of the page load energy consumption, by an average increase of 35\%, while only one extension decreases the page load energy consumption by a large effect size.
%Visible differences in the page load energy consumption are also evident in 10.16\% of the extensions (i.e., 10) with an increase of 8.69\% in small effect size, 11.48\% (i.e., 7 extensions) with an increase of 18.51\% in medium effect size, and 31.15\% (i.e., 19 extensions) with an increase of 34.58\% in large effect size.
Browser developers may provide more information about the energy consumption of the browser and the extensions to help users make informed decisions when using an extension and to motivate extension developers to improve the energy efficiency of their extensions.

%\textbf{The utilization of extensions leads to an increase in overall stabilized energy consumption by 1.44\%.}
\textbf{Even after the web page has been loaded, 45 (63\%) extensions still lead to a statistically significant increase in energy consumption. }%, when the browser is stabilized, with an average increase of 1.73\%.}
The majority of extensions, 71\% (i.e., 51 extensions), demonstrate a statistically significant difference in the stabilized energy consumption. 
63\% (i.e., 45) of the extensions consume an extra 2.0\% of the stabilized energy on average, while 8.3\% (i.e., 6) of the extensions reduce the stabilized energy consumption by an average of 1.6\%.
% \heng{Only focus on the large negative impact, similar to the previous two paragraphs}
In particular, 
% 22.22\% of extensions (i.e., 16) display small differences in the stabilized energy consumption, with an energy difference of 0.61\%, while 16.67\% (i.e., 12 extensions) exhibit medium differences, with an energy difference of 0.79\%. Furthermore,
32\% (i.e., 23) of extensions demonstrate large differences in the stabilized energy consumption, with an energy difference of 3.0\%.
Compared to the impact on the load time energy consumption, the extensions' impact on the stabilized energy consumption is relatively small. This may be explained by the fact that the extensions are less active after the pages have been loaded.

% \begin{tcolorbox}
% \noindent
\noindent\textbf{\underline{Summary and implication}:} 
Our results indicate that  %in the most typical activation mode of extensions (i.e., fully loaded), 
the use of extensions under the fully-loaded mode can lead to a statistically significant impact on the browser performance, with the largest negative impact on the load time energy consumption. 
Our observations suggest that browser and extension developers should pay attention to the performance impact of browser extensions. For example, they may provide warnings and information about the performance impact of the extensions, suggest users disable extensions in performance-critical scenarios and optimize the extensions or browser-extension integration in terms of performance. %, particularly in regard to energy efficiency.  %increase in the page load time by \textbf{18.07\%}, the page load energy consumption by \textbf{17.12\%}, and the  stabilized energy consumption by \textbf{1.73\%}.

\subsection{ RQ2: How do the activation modes of extensions affect browser performance?}\label{rq2}

\textbf{\underline{Motivation:}} In RQ1, we have analyzed the effect of the extensions on the browser performance in the expected scenario (i.e., the fully-loaded mode). 
Nonetheless, users may not always interact with extensions in the intended manner. 
For instance, a user may ignore the login process of an extension to utilize the extension. % on websites.
% for which it is not specifically tailored \heng{``no login'' and ``not tailored'' are two different things, right? Then use ``or'' or just keep one example}.
The impact of the extensions under different activation modes on browser performance remains uncertain.
Therefore, it is important to understand how different activation modes of extensions could lead to varying performance impacts
% RQ1 analyzed the overall effect of the extension on browser performance in the most typically used activation mode of extensions, known as the fully loaded mode. From a different perspective, 
% It is possible that different activation modes of extensions may result in various performance impacts.
%By understanding the performance impact with respect to different aspects of activation modes of extensions, we can gain insights into 
% Consequently, the motivation behind this RQ aims to investigate the influence of various activation modes of extensions on browser performance.
% \bihui{By addressing this RQ, we contribute to enhancing the understanding of how activation modes of extensions impact browser performance.}
and to provide insights for browser users to select and configure extensions. 

% This understanding can assist researchers and industries in being aware of potential additional energy expenses.
% Therefore, the goal behind this RQ is to investigate the impact of various extensions' activation modes on the overall browser performance of the extension.

\noindent\textbf{\underline{Approach:}}
The activation level of an extension depends on how the user configures and uses it (e.g., whether the extension is logged in or used with the designated websites). We first provide a classification of the different activation modes of the extensions, then perform experiments to test the performance impact of the different activation modes. 
% how do we define these extensions in different modes?
% \heng{Briefly describe how the classification is done}

\noindent\textbf{Classifying and testing the activation modes of extensions.}
Based on the requirements for extension access and execution (e.g., requirements for login or getting access to a website), as well as their intended usage scenarios (e.g., used with intended or designated websites), we classify activation modes of extensions into six categories: \textit{extension-free} (the baseline), \textit{fully-loaded}, \textit{no-login}, \textit{no-grant}, \textit{non-designated}, and \textit{fully-inactive}, as outlined in Table~\ref{tab:user mode}. 
\bihui{Browsing web pages without the extension installed is the extension-free mode, which serves as the baseline for performance comparison. 
The activation mode of extensions can be classified in four modes, considering whether the extension is logged in, granted access to the website, tested with the designated webpages, and enabled. 
The description of different activation modes is listed in Table \ref{tab:user mode}.
Except for the \textit{fully-loaded} mode, not all extensions necessitate certain conditions (e.g., login): activation modes of extensions are applicable only to partial extensions.
Specifically, all the 72 extensions can be tested in the \textit{fully-loaded} mode. 
13 extensions can be executed without login (i.e., \textit{no-login} mode), and 17 extensions are tested in both the websites that are not designated for the extensions (\textit{non-designated} mode) and in the websites such that the permission to access a webpage is not granted (i.e., \textit{fully-inactive} mode), respectively.}

We evaluate the performance impact of the different activation modes by comparing the measures of extensions in partial activation modes with both the baseline (the extension-free mode) and the measurement results of extensions in the fully-loaded mode.  % with five activation modes, including: "fully loaded" and four particular activation modes (i.e., "no grant", "no login", "non-designated", and "fully inactive").
% \heng{recap the fully loaded mode here too} 
%\bihui{Recalling the observations in RQ1, we can see that fully loaded extensions have a statistically significant impact on the browser performance, with the largest adverse impact on the load time energy consumption.}
%Therefore, it is recommended that users log in to the extension when prompted to do so.
%Under the fully loaded mode, an extension is given the full access (i.e., logged in and access permission granted when required, used for designated websites when needed). 
All 72 extensions are selected for testing under the fully-loaded mode (same as in RQ1).
% (i.e., logged in and access permission granted when required, used for designated websites when needed)
% However, for the other four partial activation modes, we need to select the extensions that satisfy the corresponding requirements.
We select the extensions to be tested in the partially loaded modes based on their requirements and usage scenarios. 
We test the extensions that require access permission to the tested webpages in the no-grant mode. 
We test the extensions that require login in the no-login mode.
The non-designated mode involves testing the inactive extensions that only work for designated websites and is tested with the context of generic testing scenario websites. 
% in the context of generic testing scenario websites rather than the designated websites.
The performance of the inactive extensions is assessed in the fully-inactive mode which does not log in to the extension, not grant access to the webpage, and is tested with the generic testing scenario websites rather than the designated websites to the extensions that only work for the designated websites.
% \heng{not clear, be precise: you only test the extensions that require login, grant to access websites, and only work for designated websites, right?-\bihui{We also test the fully inactive mode.}}.
We measure a total of 40 extensions across the partial activation modes. 
\bihui{Tables \ref{tab:mode-result-free} and \ref{tab:mode-result-full} list the distribution of the number of extensions tested in each activation mode. As shown in Tables \ref{tab:mode-result-free} and \ref{tab:mode-result-full}, we can compare the extension-free mode with the fully-loaded mode.}
% \heng{why not put these numbers in Table 3? -\bihui{because we want make Table 4 be consistent with Table 2. Moreover, put these numbers in Table 4 is easier for audiance to compare \# of changes with the total number of extensions}}.
% We compare the measurement results of extensions in the fully loaded mode with the measures of extensions in partial activation modes \heng{merge this with the previous sentence about comparing with the baseline}.
% we report the results of the fully loaded model for the corresponding extensions used for each partial activation mode.
% , with 10 extensions in the mode of no access permission, 8 in the no login mode, 11 in the inactive mode, and 11 in the fully inactive mode.
We follow the same measurement procedures, as outlined in Section \ref{measurement procedure}.
% \heng{Missing details on how the tests are run (what websites, run how may times etc.).}
The measurements are recorded and then normalized using the procedure outlined in Section \ref{normalization}. 
% The normalization is performed relative to the measurements in the "extension-free" mode. 
% Measurements are then recorded and then normalized following the normalization code in Section \ref{normalization}. 
% The normalization is conducted with respect to the measurements in the extension-free mode. 

\begin{table}
\Large
% \vspace{-2mm}
\centering
\caption{Activation modes of extensions. }
% \heng{What if an extension does not need login/grant? Use table notes to explain that when login or grant are not required, we always consider them as logged or granted. -\bihui{For the extensions that does not need login/grant themselves, no test is performed in no grant/no login modes. However, if they are only work for designated websites, they are still tested in full inactive mode.}} \heng{maybe add the number of extensions in the first column (e.g., Fully loaded\\(72)) -\bihui{but we already present the number of extensions in Table 4}}}
% \vspace{-2mm}
\label{tab:user mode}
\resizebox{\linewidth}{!}{
\begin{threeparttable}
\begin{tabular}{l|ccccc|l} 
\hline
\multicolumn{1}{c|}{\begin{tabular}[c]{@{}c@{}}Mode\\name\end{tabular}} & Login\tnote{*} & Grant\tnote{*} & \begin{tabular}[c]{@{}c@{}}Designated\\webpage\tnote{**}\end{tabular} & \begin{tabular}[c]{@{}c@{}}Extension\\installed\end{tabular} & \begin{tabular}[c]{@{}l@{}}Extension\\enabled\end{tabular}  & \multicolumn{1}{c}{Description}                                                                                                                                \\ 
\hline
Extension-free                                                          & 0     & 0     & 0                                                       & 0                                                            & -                      & \begin{tabular}[c]{@{}l@{}}No extensions are installed. \textbf{(Baseline)}\end{tabular}                                                           \\ 
\hline
Fully-loaded                                                            & 1     & 1     & 1                                                       & 1                                                            & 1                      & \begin{tabular}[c]{@{}l@{}}Extension is logged in; extension is\\tested with the~designated~webpages;\\extension's access permission to\\the webpages is granted \textbf{(Baseline)}\end{tabular}         \\ 
\hline
No-login                                                                & 0     & 1     & 1                                                       & 1                                                            & 1                      & An extension is not logged in                                                                                                                          \\ 
\hline
No-grant                                                                & 1     & 0     & 1                                                       & 1                                                            & 1                      & \begin{tabular}[c]{@{}l@{}}Extension's access permission to the\\tested webpages is not granted\end{tabular}                                                            \\ 
\hline
Non-designated                                                                & 1     & 1     & 0                                                       & 1                                                            & 0 \tnote{***}                     & \begin{tabular}[c]{@{}l@{}}An extension is not tested with the~\\designated~webpages\end{tabular}                                                                 \\ 
\hline
Fully-inactive                                                          & 0     & 0     & 0                                                       & 1                                                            & 0                      & \begin{tabular}[c]{@{}l@{}}Access to an extension and access \\permission to a webpage are not granted;\\ not tested with the~designated~webpages\end{tabular}  \\
\hline
\end{tabular}
\begin{tablenotes}
      \Large
      \item 1 represents True (i.e., performed); 0 represents False (i.e., not performed); ``-'' represents not applicable.

      \item * If an extension does not need to be logged in or to be granted access, it is always considered as being logged in or granted access.

      \item ** A designated webpage of an extension is the webpage that an extension is designed for (e.g., GitHub); the generic extensions (corresponding to the generic scenario in Table~\ref{tab:scenario}) have all webpages as their designated ones. 
      
      \item *** Extensions are naturally disabled when accessing non-designated webpages.

\end{tablenotes}
\end{threeparttable}
}
% \vspace{-8mm}
\end{table}

\noindent\textbf{Statistical analysis of the experiment results.}
% \heng{If it is done in the same way as in RQ1, just simply say we use the same method as in RQ1 to ... Then remove all the details here} 
Similar to RQ1, we apply the Wilcoxon signed-rank test and the Cliff's $\delta$ test to determine the degree of difference of the measurements between running an activation mode of an extension and running in the extension-free, as well as between running an activation mode of an extension and the fully-loaded mode. 
We test the following two hypotheses:

$H0_2$: \textit{there is no difference in the distributions of performance metric values between the paired observations in  a partial activation mode and the \textbf{extension-free mode}.}

$H0_3$: \textit{there is no difference in the distributions of performance metric values between the paired observations in a partial activation mode and the \textbf{fully-loaded mode}.}

%The measurements that pass the statistical tests are considered to have a statistically significant difference and be practically meaningful. 
%\footnote{\url{https://github.com/anon-author-research/suppmaterial-impact-of-extensions-on-browser-performance}}.
 % The effect sizes are reported in our replication package.
% Similar to RQ1, the Wilcoxon signed-rank test carrying on a confidence level of 5\% are applied to compare two normalized attributes (e.g., the energy consumption of payload) taken with and without the extension enabled. 
% Subsequently, we use the Cliff’s Delta test to quantitatively determine the degree of difference in the effect size between the paired attributes.

\noindent\textbf{Performance change ratio.} %Therefore, we use such measurements to yield the 
When an extension leads to a statistically significant performance difference between the performance  metrics under an activation mode and the extension-free mode or the fully-loaded mode, we calculate the change ratio of the performance metric, 
%change ratio for each performance metric across the activation modes of extensions, 
following the same method described in RQ1.
%The overall change ratio, as shown in Table \ref{tab:mode-result}, presents the mean of the change ratios across all effect sizes.

\noindent\textbf{\underline{Findings:}}
% \heng{The discussion of this RQ's results need significant re-organization. Suggested flow: First paragraph remains (all activation modes have impact). Second paragraph: even when extensions are not used in the expected manner (i.e., when access to the extension is not granted or a user is not logged in), it still leads to significant energy consumption. Third paragraph: even when extensions are inactive (not used for target websites or fully deactivated), they can still lead to significant energy consumption and increase response time. Finally, we can discuss how these modes differ from the fully loaded modes (we may need a small table to show the relative difference or combine them with Table 6/7 (after moving effect size things to replicaiton package).)}
\textbf{Browser performance is impacted by the use of extensions regardless of their activation modes.}
The results of testing $H0_2$ and $H0_3$ are presented in Tables \ref{tab:mode-result-free} and \ref{tab:mode-result-full}, indicating the relative performance impact of the extensions in their four partial activation modes, namely: no-grant, no-login, non-designated, and fully-inactive, in comparison to the two baselines (the extension-free mode and the fully-loaded mode). 
% \heng{shall we call the fully loaded mode as a baseline too? then need to indicate it in Table 3 and the related text -\bihui{I think if we call the fully loaded mode as a baseline, it would be confusing for readers to distinguish the extension-free and fully-loaded modes when reading it.}}. 
%\bj{
%The results for $H0_3$ are indicated by the figures enclosed in brackets in Table \ref{tab:mode-result}.}
Table \ref{tab:mode-result-free} shows that all partial activation modes of the extensions can statistically significantly impact the browser performance in terms of the studied performance metrics.
%These performance changes are reflected in the page load time, varying from -6.51\% to 80.27\%, the page load energy consumption, ranging from 11.53\% to 14.82\%, and the stabilized energy consumption, varying from -0.47\% to 8.79\%. 
%In particular, the fully inactive mode has the most significant impact on the browser performance, especially on the page load time and page load energy consumption.  
%This may be because the extensions make attempts to gain access when a webpage is loading, which may cause significant extra overhead.
Overall, 12 out of the 13 extensions tested with the no-grant mode, 16 out of the 17 extensions tested with the no-login mode, 8 out of the 11 extensions tested with the non-designated mode, and 11 out of the 11 extensions tested with the full-inactive mode statistically significantly impact at least one of the studied performance metrics.

\textbf{Using extensions in unexpected circumstances (i.e., when access to a webpage is not granted or an extension is not logged in) can still lead to impaired browser performance, especially more energy consumption.} 
For example, 38\% of the tested extensions (i.e., 5) running without a granted permission (i.e., no-grant mode) and 72\% of the extensions (i.e., 12) tested without the required login (i.e., no-login mode) lead to an average of 19\% and 18\% increase in the page load energy consumption, respectively.
% \heng{this kind of statements are not accurate, you cannot say varying from ... as they are averages. Instead, you should say ``leading to an average of 14.82\% and 11.53\% increase in the page load energy consumption, respectively''} \heng{fix all such statements in the rest of the RQ!}
88\% of extensions (i.e., 15) in the no-login mode and 85\% of extensions (i.e., 11) in the no-grant mode result in a slight increase in the stabilized energy consumption, leading to an average of 2.8\% and 2.9\% increase, respectively.
When users opt to disregard the login process of the extensions, it is observed that approximately 41\% of the tested extensions (i.e., 7)  significantly contribute to the page load time, leading to an average delay of 33\% on average.
In comparison, denying access permission to the website is more likely to decrease the response time while leading to an increase in energy consumption during the page load and the stabilized periods.
In particular, 46\% of the extensions (i.e., 6)  significantly impact the page load time, resulting in an average of 6.3\% decrease on average.
%  the response time of payload decreases by 6.51\% by 50.00\% of the extensions (i.e., 5) when the access permission to the website is not granted.
% In comparison to the 1.44\% change ratio observed in the fully loaded mode, 90.00\% of the extensions (i.e., 9) in the no grant mode and 100.00\% of the extensions (i.e., 8) in no login mode slightly increase the stabilized energy consumption by 2.72\% to 2.81\%.
% \heng{Briefly explain why the unexpected circumstances lead to performance impact.}
Under unexpected circumstances, extensions may not be able to utilize their full functionality to efficiently load and interact with web content, leading to suboptimal performance and potential disruptions during browsing activities.
% \textbf{When access permission to the website is not granted, it may decrease the response time while leading to increase in energy consumption during the payload and the stabilized periods.}
% In particular, the response time of payload decreases by 6.51\% by 50.00\% of the extensions (i.e., 5) when the access permission to the website is not granted. However, the energy consumption during the payload and the stabilized periods are increased 14.82\% and 2.81\%, respectively.

% \textbf{\heng{be more rigorous: this claim only hold for the five extensions that have significant impact. We may use bold sentences like this: When access to an extension is not granted, it may decrease the response time while leading to increase in energy consumption during the payload and the stabilized periods.} The response time of payload decreases by 6.51\% in overall when the access permission to the website is not granted. When the user does not use the extension properly, such as in the fully inactive mode, the response time of payload soars by 80.27\%.} 25\% of the extensions (i.e., 2) result in an increase of the response time of payload by 13.04\%. When the extension is deployed in the fully loaded mode, the response time is only affected by 1.00\%. 
% The changes of the response time of payload indicate that any changes in the activation mode of extensions applied to the extension result in a change in the response time of payload.
% \input{tables/rq2.tex}
% \usepackage{graphicx}
% \usepackage{multirow}

\begin{table}
\Large
\vspace{-2mm}
\centering
\caption{Comparing the performance impact of different activation modes of the extensions with the extension-free mode.} 
\vspace{-2mm}
\label{tab:mode-result-free}

\resizebox{\linewidth}{!}{%
\begin{tabular}{l|c|cr|cr|cr|cr} 
\hline
\multicolumn{2}{c|}{Mode} & \multicolumn{2}{c|}{No-grant} & \multicolumn{2}{c|}{No-login} & \multicolumn{2}{c|}{Non-designated} & \multicolumn{2}{c}{Fully-inactive} \\ 
\hline
\begin{tabular}[c]{@{}l@{}}Performance\\ metrics\end{tabular} & Tendency & \multicolumn{1}{l}{Count} & \multicolumn{1}{c|}{Ratio} & \multicolumn{1}{l}{Count} & \multicolumn{1}{c|}{Ratio} & \multicolumn{1}{l}{Count} & \multicolumn{1}{c|}{Ratio} & \multicolumn{1}{l}{Count} & \multicolumn{1}{c}{Ratio} \\ 
\hline
\multirow{3}{*}{\begin{tabular}[c]{@{}l@{}}Page load\\time\end{tabular}} & Increase & 2 & 43\% & 5 & 46\% & 1 & 335\% & 4 & 231\% \\ 
\cline{2-2}
 & Decrease & 4 & -6.9\% & 2 & -4.8\% & 0 & \multicolumn{1}{c|}{-} & 3 & -36\% \\ 
\cline{2-2}
 & Overall & 6 & -6.3\% & 7 & 33\% & 1 & 335\% & 7 & 80\% \\ 
\hline
\multirow{3}{*}{\begin{tabular}[c]{@{}l@{}}Page load\\energy\\consumption\end{tabular}} & Increase & 5 & 19\% & 12 & 18\% & 5 & 14\% & 4 & 577\% \\ 
\cline{2-2}
 & Decrease & 0 & \multicolumn{1}{c|}{-} & 0 & \multicolumn{1}{c|}{-} & 0 & \multicolumn{1}{c|}{-} & 3 & -29\% \\ 
\cline{2-2}
 & Overall & 5 & 19\% & 12 & 18\% & 5 & 14\% & 7 & 12\% \\ 
\hline
\multirow{3}{*}{\begin{tabular}[c]{@{}l@{}}Stabilized \\energy\\consumption\end{tabular}} & Increase & 11 & 2.8\% & 15 & 2.9\% & 3 & 1.6\% & 8 & 13\% \\ 
\cline{2-2}
 & Decrease & 0 & \multicolumn{1}{c|}{-} & 0 & \multicolumn{1}{c|}{-} & 5 & -0.79\% & 3 & -28\% \\ 
\cline{2-2}
 & Overall & 11 & 2.8\% & 15 & 2.9\% & 8 & -0.47\% & 11 & 8.8\% \\ 
\hline
\multicolumn{2}{l|}{\# of extensions being tested} & \multicolumn{2}{c|}{13} & \multicolumn{2}{c|}{17} & \multicolumn{2}{c|}{11} & \multicolumn{2}{c}{11} \\
\hline
\end{tabular}
}
\begin{tablenotes}
      \small
      \item We share our Cliff's delta results in our replication package.
\end{tablenotes}
\end{table}

\begin{table}
\Large
\vspace{-2mm}
\centering
\caption{Comparing the performance impact of different activation modes of the extensions with the fully-loaded mode.}

\vspace{-2mm}
\label{tab:mode-result-full}
\resizebox{\linewidth}{!}{%
\begin{tabular}{l|c|cr|cr|cc|cr} 
\hline
\multicolumn{2}{c|}{Mode} & \multicolumn{2}{c|}{No-grant} & \multicolumn{2}{c|}{No-login} & \multicolumn{2}{c|}{Non-designated} & \multicolumn{2}{c}{Fully-inactive} \\ 
\hline
\begin{tabular}[c]{@{}l@{}}Performance\\ metrics\end{tabular} & Tendency & \multicolumn{1}{l}{Count} & \multicolumn{1}{c|}{Ratio} & \multicolumn{1}{l}{Count} & \multicolumn{1}{c|}{Ratio} & \multicolumn{1}{l}{Count} & Ratio & \multicolumn{1}{l}{Count} & \multicolumn{1}{c}{Ratio} \\ 
\hline
\multirow{3}{*}{\begin{tabular}[c]{@{}l@{}}Page load\\time\end{tabular}} & Increase & 1 & 17\% & 3 & 17\% & 0 & - & 4 & 21\% \\ 
\cline{2-2}
 & Decrease & 3 & -3.1\% & 2 & -7.4\% & 0 & - & 2 & -41\% \\ 
\cline{2-2}
 & Overall & 4 & -2.1\% & 5 & 7.0\% & 0 & - & 6 & 11\% \\ 
\hline
\multirow{3}{*}{\begin{tabular}[c]{@{}l@{}}Page load\\energy\\consumption\end{tabular}} & Increase & 1 & 44\% & 9 & 19\% & 3 & \multicolumn{1}{r|}{11\%} & 6 & 14\% \\ 
\cline{2-2}
 & Decrease & 0 & \multicolumn{1}{c|}{-} & 0 & \multicolumn{1}{c|}{-} & 1 & \multicolumn{1}{r|}{-61\%} & 1 & -5.9\% \\ 
\cline{2-2}
 & Overall & 1 & 44\% & 9 & 19\% & 4 & \multicolumn{1}{r|}{7.9\%} & 7 & 11\% \\ 
\hline
\multirow{3}{*}{\begin{tabular}[c]{@{}l@{}}Stabilized \\energy\\consumption\end{tabular}} & Increase & 7 & 4.0\% & 9 & 1.6\% & 7 & \multicolumn{1}{r|}{1.3\%} & 9 & 1.3\% \\ 
\cline{2-2}
 & Decrease & 0 & \multicolumn{1}{c|}{-} & 1 & -0.83\% & 0 & - & 1 & -3.8\% \\ 
\cline{2-2}
 & Overall & 7 & 4.0\% & 10 & 1.6\% & 7 & \multicolumn{1}{r|}{1.3\%} & 10 & 1.3\% \\ 
\hline
\multicolumn{2}{l|}{\# of extensions being tested} & \multicolumn{2}{c|}{13} & \multicolumn{2}{c|}{17} & \multicolumn{2}{c|}{11} & \multicolumn{2}{c}{11} \\
\hline
\end{tabular}
}
\begin{tablenotes}
      \small
      \item We share our Cliff's delta results in our replication package.
\end{tablenotes}
\vspace{-4mm}
\end{table}
\textbf{Even when extensions are not used for their designated websites (e.g., TubeBuddy for {\tt Videos}) or are fully deactivated, they can still lead to significant energy consumption and increase the page load time.} 
In particular, 7 to 11 (64\% to 100\%) of the extensions tested under the fully-inactive mode significantly impact the studied performance metrics. %the fully inactive mode results in statistically significant changes in all the performance metrics, 
% leading to an 80.27\% increase in the page load time, an 11.69\% increase in the page load energy consumption, and an 8.79\% increase in the stabilized energy consumption, on average. \heng{briefly explain why}
Such observation may be attributed to the fact that extensions may still run background processes and built-in functionalities or scripts, consuming resources even when they are not actively being used.
The non-designated mode results in statistically significant changes on 5 extensions (45\%) in the page load energy consumption by an average of 14\% on average.
% \heng{be accurate in wording: when focusing on the extensions that increase the consumption: XX extensions significantly increase the page load energy consumption by 14.46\% on average; when focusing on all extensions that either increase/decrease consumption: YY extensions significantly impact the page load energy consumption, leading to an average of 14.46\% increase on average}. % and the stabilized energy consumption by -0.47\% on average.
The negative performance impact of the non-designated modes may be because the extensions make attempts to  access the web content when a webpage is loading, which may cause significant extra overhead.

% \textbf{The energy consumption of payload remains relatively stable across different activation modes of extensions, as evidenced by a minimal change of 11.53\% compared to the fully loaded mode.} Effect size is all above small, in particular, 3 out of 8 the effective size are large. Small effect size suggests that the results have limited practical applications and large effect size indicates that the results have practical significance.
% 40\% of the extensions (i.e., 4) in the no grant mode, 87.5\% of the extensions (i.e., 7) in no login mode, 45.45\% of the extensions (i.e., 5) in inactive mode, and 63.64\% of the extensions (i.e., 7) in fully inactive modes exhibit a slightly increase in the energy consumption of payload, with variations ranging from 0.02\% to 3.25\%.

% \heng{Finally, we can discuss how these modes differ from the fully loaded modes (we may need a small table to show the relative difference or combine them with Table 6/7 (after moving effect size things to replication package).)}
% \heng{update after Table 4 updated}
%\textbf{The page load time and the stabilized energy consumption tend to be higher compared to the fully loaded mode, whereas the page load energy consumption is generally lower.}
\textbf{The partial activation modes of extensions can lead to even worse performance impact than the fully-loaded mode.} 
Table~\ref{tab:mode-result-full} indicates the relative performance impact of the different activation modes in comparison to the fully-loaded mode. Surprisingly, all partial activation modes can lead to a worse performance impact even than the fully-loaded mode. For example, 9 out of the 17 extensions tested under the no-login mode exhibit a statistically significant increase in the page load energy consumption in comparison to the fully-loaded mode, leading to an average increase of 19\% in the metric value. 4 out of the 11 extensions tested under the fully-inactive mode increase the page load time over the fully-loaded mode, with an average increase of 21\%.
When extensions are not used in the intended mode (i.e., fully-loaded), %such as non-designated and fully inactive modes, even though the extensions are not deployed on the designated website, 
they still run and consume resources (e.g., through continuous attempts of obtaining access), which may consume more resources and lead to worse user experience than the fully-loaded mode. %, which aligns with our observation that most extensions have a negative impact on browser performance. Furthermore, this deterioration is exacerbated when extensions are used improperly, specifically in the fully inactive mode.
The surprising results indicate the need for better performance testing of the extensions under the unintended usage scenarios.

% \vspace{-2mm}
% \begin{tcolorbox}

% \noindent
\noindent\textbf{\underline{Summary and implication}:} 
%Our results indicate that any changes in the activation mode of extensions can affect the browser performance, ranging from \textbf{-6.51\% to 80.27\%} in the page load time, \textbf{11.53\% to 14.82\%} in the page load energy consumption, and \textbf{-0.47\% to 8.79\%} in the stabilized energy consumption.
We observe that browser performance can be negatively impacted by the use of extensions even when they are used in unexpected circumstances (e.g., not logged in or access to a webpage not granted) or are not active (e.g., not used for designated websites).
Surprisingly, unintended usage scenarios of extensions can even lead to worse performance impact than the fully-loaded mode.
Extension users should be aware of such performance impact to optimize their configurations of the extensions (e.g., avoiding improper use). Our findings also suggest that browser and extension developers should take action on reducing the performance impact of extensions under unintended usage scenarios (e.g., better performance testing and optimization for such scenarios). %when the extensions work in a non-designated mode.
% \end{tcolorbox}
% \vspace{-2mm}
\subsection{ RQ3: What factors of extensions influence browser performance?}\label{rq3} 

\textbf{\underline{Motivation:}}
Browser extensions are designed with various characteristics, such as the adopted privacy practices and activation modes to provide users with various functionalities. % measured in lines of code, 
%to suit users' demands. 
% Nevertheless, some design decisions may have a hidden impact on the browser performance.
In RQ2, we have observed that different activation modes of extensions lead to different impacts on the browser performance. 
\bihui{In this RQ, we are interested in understanding multifaceted factors that can significantly influence browser performance. 
We want to unravel the interplay of these factors.
%\bj{Uncovering the correlation between these factors and the  performance metrics}
% it helps us delve into the hidden variables that affect the page loading time or are more energy demanding. 
With such insights, extension developers can use the knowledge of performance influencing factors to optimize their extensions.
Moreover, extension users can make more informed decisions to select extensions that best meet their needs by considering various factors that may potentially affect the browser performance.}
%practitioners to select extensions and highlights the influential factors for practitioners to optimize the extension's performance further.

% \begin{figure}
% \centering
%     \includegraphics[width=0.48\textwidth]{spearman.png}
%     % \includegraphics[scale=0.7]{spearman.png}
%     \caption{Hierarchical clustering of extension variables according to Spearman’s $|\rho|$ \heng{move this to repilicaiton package}}
%     \label{fig:spearman}
% \end{figure}
\noindent\textbf{\underline{Approach:}}
We extract a set of factors (e.g., code metrics) of the extensions and leverage a statistical model to understand the influential factors.

\noindent{\textbf{Extracting extension factors.}}
% We considered different types of factors of the extensions, such as code metrics and file characteristics, to understand their influence on browser performance.
Table~\ref{tab:factors} lists 105 considered factors of the extensions.
\bj{We consider the common factors that might impact performance from  five aspects, including code metrics, file characteristics, privacy practices, user perspectives, and activation modes of extensions.}
We employ SciTools' Understand\footnote{\url{https://scitools.com/}} to analyze the inlining factors of the extensions.
The Understand tool evaluates the code using various code metrics\footnote{\url{https://documentation.scitools.com/pdf/metricsdoc.pdf}} and also provides entity information, such as the number of classes and public methods utilized in the extension. 
We consider the code metrics proposed by \citet{ck} and \citet{lz}.
Beyond the code metrics and entities, we take into account the types of collected data and file characteristics (e.g., the file size in different kinds of files related to the extension, such as the size of .png typed files).
72 compressed extension projects (i.e., .crx files) are extracted via the online web tool\footnote{\url{https://crxextractor.com/}} to obtain the source code.
Besides inlining factors, we complement additional factors as per the inherent nature of the extensions, including the belonged categories, user perspective, and usage scenarios (i.e., activation modes).
A total of 105 potential factors of the extensions that may impact the browser performance are listed in Table \ref{tab:factors}.
%With the help of the scientific tool, 
%this meticulous selection of factors aims to affirm the reliability of the identified aspects.

To understand the interplay of these factors on the browser performance, we construct a statistical model.
The dataset is collected from the same experiment performed in Section \ref{rq2}, and the characteristics of the extensions (e.g., extension category) are collected based on the steps described in Section \ref{CE}.
% We \bihui{manually} identify 17 potential factors of extensions that may impact the browser performance, as listed in Table \ref{tab:factors} (i.e., the ``fixed variables'').
% These factors include privacy practices adopted 
% % \heng{use ``adopted'' instead of ``supported'' (the privacy practices are used to collect data, not supporting user needs)}
% by the extensions, their sizes, popularity (in terms of the number of users, raters, and rating scores), as well as their usage scenarios (i.e., activation modes). 
% Based on these factors, we construct a statistical model to understand the ones that significantly influence the browser performance.
%  Correlations may exist among these factors, thus we perform correlation analysis and redundancy analysis to remove the high correlations before constructing the model.
% The identities of the tested websites are used as the control variable (i.e., the ``random variable'').
% % \heng{change ``150 websites'' to ``Website used for test''? Only one website is used for each test case, right? Then change ``Websites'' to ``The website'' in the description.}
% %\heng{Missing some information on how the data for training/testing the model is collected. If the data collection is the same as an earlier experiment, then briefly explain that -%\bihui{plz check below}}
% We collect the dataset from the same experiment performed in RQ2.
% The characteristics of the extensions (e.g., extension category) are collected based on the steps described in Section \ref{CE}.

\begin{center}
\setlength\LTleft{-8mm}
{\normalsize\tabcolsep=3pt
\begin{longtable}{l|l|l}
\caption{Studied factors that may influence the performance impact of browser extensions. }
% \heng{The file size metrics can be summarized as one item ``Sizes of different type of files''. The complete list could be put in the replication package \bihui{-solved}}}
\label{tab:factors}\\

% \resizebox{\linewidth}{!}{
% \vspace{-2mm}
% \begin{threeparttable}

% \begin{tabular}{l|l|l} 
% \hline
\toprule
\multicolumn{1}{c|}{Variable Type} & \multicolumn{1}{c|}{Potential Influential Factors} & \multicolumn{1}{c}{Description}  \\ \hline
\endfirsthead

\multicolumn{3}{c}%
{{ \tablename\ \thetable{} -- continued from previous page}} \\
\hline \multicolumn{1}{c|}{Variable Type} & \multicolumn{1}{c|}{Potential Influential Factors} & \multicolumn{1}{c}{Description} \\ \hline 
\endhead

\hline \multicolumn{3}{|r|}{{Continued on next page}} \\ \hline
\endfoot

\hline \hline
\endlastfoot
% Variable Type & Potential Influential Factors & \multicolumn{1}{c}{Description} \\ 
% \hline
\multicolumn{1}{l|}{} & Ambient Module & Number of ambient modules \\
 & Class & Number of classes~ \\
 & ClassFunction & Number of class functions \\
 & Enum & Number of enum classes \\
 & File & Number of files \\
 & Interface & Number of interfaces \\
 & Method & Number of methods \\
 & Namespace & Number of namespaces \\
 & Private Method & Number of private methods \\
 & Public Method & Number of public methods \\
 & Public Static Method & Number of public static methods \\
 & Function & Number of functions \\
 & Unnamed Function & Number of unnamed functions \\
 & WMC (Weighted Methods per Class) & Number of local (not inherited) methods \\
 & DIT (Depth of Inheritance Tree) & Maximum depth of class in inheritance tree \\
 & ev(G) (Essential complexity) & \begin{tabular}[c]{@{}l@{}}The number of decision points + 1 \\after control graph reduction\end{tabular} \\
Code Metric & CC (Cyclomatic Complexity) & The number of decision points + 1 \\
 & NPATH (Number of Possible Paths) & \begin{tabular}[c]{@{}l@{}}Number of unique paths trhough a body of code,\\~not countingabnormal exits or gotos\end{tabular} \\
 & CLOC (Comment Lines of Code) & Number of lines containing comment \\
 & LOC (Lines of Code) & Number of lines containing source code \\
 & BLOC (Blank Lines of Code) & Number of blank lines \\
 & NL (Number of Lines) & Number of Lines \\
 & NPM (Number of Public Methods) & Number of local (not inherited) public methods \\
 & NPRM (Number Private Methods) & Number of local (not inherited) private methods \\
 & RFC (Response for a Class) & Number of methods, including inherited ones \\
 & NIV (Number of Instance Variables) & Number of instance variables \\
 & NIM (Number of Instance Methods) & Number of instance methods \\
 & NV (Number of Variables) & Number of class variables \\
 & NOC (Number of Children) & Number of immediate subclasses \\
 & IFANIN & Number of immediate base classes \\ 
\hline
\begin{tabular}[c]{@{}l@{}}File \\Characteristic\end{tabular} & Sizes of different type of~files & Size of various types of files in the extension \\ 
\hline
\multirow{8}{*}{\begin{tabular}[c]{@{}l@{}}Privacy\\Practice\end{tabular}} & Total \# of Privacy Practices Used & \begin{tabular}[c]{@{}l@{}}Number of privacy practice properties\\adopted by the extension\end{tabular} \\ 
\cline{2-3}
 & Location & \multirow{7}{*}{\begin{tabular}[c]{@{}l@{}}Seven privacy practice items\\adopted by the extension\end{tabular}} \\
 & User Activity &  \\
 & Website Content &  \\
 & Web History &  \\
 & P.I.I. &  \\
 & Authentication Information &  \\
 & Personal Communications &  \\ 
\hline
\multirow{3}{*}{\begin{tabular}[c]{@{}l@{}}User\\Perspective\end{tabular}} & Number of Users & Number of users installing the extension \\
 & Number of Raters & Number of users rating the extension \\
 & Extension Size & The extension package size \\
 & Rating Score & \begin{tabular}[c]{@{}l@{}}The rating score of the extension\\in Chrome Web Store\end{tabular} \\ 
\hline
\multirow{4}{*}{\begin{tabular}[c]{@{}l@{}}Activation\\Mode\end{tabular}} & Logged in & \multirow{4}{*}{Four activation modes of extensions} \\
 & Access to webpage granted &  \\
 & Non-designated mode &  \\
 & Fully-inactive mode &  \\
\bottomrule
\end{longtable}
\vspace{-4mm}
\begin{tablenotes}
    \small
       \item P.I.I. refers to Personally Identifiable Information
       \item The complete list is put in the replication package.
\end{tablenotes}
\vspace{-2mm}
}
\end{center}

\noindent\textbf{Redundancy analysis.} 
% \heng{Did you do correlation analysis before the redundancy analysis? If not, I think we need to provide reason here}
% The elastic net regression model is a statistical model that can handle multicollinearity by selecting a subset of predictors that are correlated \citep{ENR-correlated}, but not redundant, as redundant predictors unnecessarily deteriorate the performance of the regression models \citep{ENR-Redun1,ENR-Redun2}.
Variable selection is incorporated into the model-building procedure to aid in raising the accuracy \citep{ENR-correlated, noCorrelation}, whereas redundancy is not considered, as redundant predictors unnecessarily deteriorate the performance of the regression models \citep{ENR-Redun1,ENR-Redun2}.
To uphold model robustness and minimize complexity, we conduct a redundancy analysis before building models to avoid redundant factors that could interfere with each other using the {\tt redun} function from the {\tt Hmisc} library in R.
This analysis assesses the redundancy correlation among the 115 identified factors.
Factors that can be explained by other factors with an R$^2$  larger than 0.90 are deemed highly redundant and thus eliminated from the result set. 
As a result, 24 factors (i.e., \textit{Ambient Module}, \textit{Interface}, \textit{Namespace}, \textit{Private Method}, \textit{Enum}, \textit{Class}, \textit{Class Function}, \textit{Public Method}, \textit{Unnamed Function}, \textit{Method}, \textit{File},
\textit{NIM}, 
\textit{NPRM}, 
\textit{NPM}, 
\textit{DIT}, 
\textit{RFC}, 
\textit{ev(G)}, 
\textit{NIV}, 
\textit{NL}, 
\textit{CC}, 
\textit{IFANIN}, 
\textit{WMC}, 
\textit{CLOC}, and 
\textit{BLOC}) among the code metrics, 40 factors (i.e.,
\textit{Size of .mem, .data, .psd, .sendkeys, .datepick, .tz, .blockUI, .dat, .vtt, .otf, .mjs, .1, .6, .ts, .patch, .targ, .lock, .opts, .cjs, .in, .def, .jst, .coffee, .htm, .zip, .avif, .config, .md, .mp4, .conf, .sortable, .webm, .js, .woff2, .html, .txt, .gif, .woff, .ico, .css, and .xml files}) among the file characteristics, 
\textit{extension size}, and 
\textit{total number of privacy practices used} are redundant and not considered in our model.
In the end, we keep 37 factors after the redundancy analysis for inclusion in our model.

\noindent\textbf{Model Construction and Evaluation.} 
To study the interplay of the factors on performance metrics (e.g., the page load time), we build elastic net regression models \citep{enrm,enrm2,enrm3} using the extension factors as explanatory variables and the performance metrics as response variables.
% In addition, the elastic net regression model can flexibly produce a lean set of important variables especially in high dimensional datasets.
 We use the {\tt train} function \citep{caret} provided by the {\tt caret} library in R to construct elastic net regression models for each performance metric with the gaussian method. 
 We normalize each factor by dividing each value by the largest value in the dataset, which facilitates the convergence of coefficients.

In the quest for finely tuned hyperparameters, we utilize the {\tt trainControl} \citep{trainControl, trainControl_validation, trainControl_validation2} from the {\tt caret} package. 
We opt for adaptive resampling ({\tt adaptive\_cv}), employing adaptive cross-validation for a total of 100 times parameter tuning. 
% \heng{Explain what/which portion of data are used for the tuning process, if applicable}
We assess each performance metric with regard to extension factors in the tuning process.
This process leverages the Bradly-Terry resampling method, which manages a substantial number of tuning parameter settings, incorporated with a random search \citep{trainControl_random} to ensure comprehensive coverage.
To understand the fitness of the tuned model, we explore a range of tuning parameter combinations (i.e., {\tt tuneLength}) through a random search.
In essence, {\tt tuneLength} signifies the scope or breadth of the search space for optimal tuning parameters.
We systematically vary the {\tt tuneLength} within the range of 1 to 300, allowing for a comprehensive evaluation of model fitness across different parameter settings.
The criteria for selecting the optimal model hinges on the root-mean-square error (RMSE), which is also chosen by the function selection in fine-tuning process by default. 
The model with the lowest RMSE is targeted.
RMSE is used to select the optimal model using the smallest value.
As a result, our elastic net regression models illustrate a RMSE of 0.020 for the page load time, 0.043 for the page load energy consumption, and 0.067 for the stabilized energy consumption, respectively.

\begin{table}[!b]
% \vspace{-4mm}
\centering
\caption{The significance of the extension factors for explaining the performance metrics (based on the linear mixed-effects models). }
% \vspace{-4mm}
\label{tab:lmer-time}
\label{tab:lmer-payload-energy}
\label{tab:lmer-stabilized-energy}
\resizebox{\linewidth}{!}{%
\begin{tabular}{l|l|r|r|r} 
\cline{3-5}
\multicolumn{1}{l}{} &  & \begin{tabular}[c]{@{}l@{}}Page Load \\Time\end{tabular} & \begin{tabular}[c]{@{}l@{}}Page Load \\Energy \\Consumption\end{tabular} & \begin{tabular}[c]{@{}l@{}}Stabilized \\Energy \\Consumption\end{tabular} \\ 
\hline
Type & Measurement & \multicolumn{3}{c}{\textcolor[rgb]{0.122,0.122,0.129}{Coefficients}~} \\ 
\hline
\multirow{6}{*}{Code Metric} & Number of Functions &  &  & ~7.6e-03 \\
 & Number of Public Static Methods &  &  & ~1.5e-03 \\
 & \begin{tabular}[c]{@{}l@{}}Code complexity\\(NPATH - Number of unique paths \\through a body of code)\end{tabular} &  &  & ~11e-03 \\
 & Number of Lines of Code &  & -3.8e-03 & -14e-03 \\
 & Number of Lines &  &  &  \\
 & Number of Children/subclasses &  &  & -0.47e-03 \\ 
\hline
\multirow{17}{*}{\begin{tabular}[c]{@{}l@{}}File \\Characteristic\end{tabular}} & Size of .svg files &  &  & ~6.4e-03 \\
 & Size of .png files &  & -4.9e-03 & -13e-03 \\
 & Size of .eot files &  & -1.3e-03 & -13e-03 \\
 & Size of .json files &  &  & ~2.3e-03 \\
 & Size of .ttf files &  &  &  \\
 & Size of .ogg files &  &  & ~3.4e-03 \\
 & Size of .map files &  &  &  \\
 & Size of .jpg files &  &  &  \\
 & Size of .bak files &  &  & -4.5e-03 \\
 & Size of .mp3 files &  &  &  \\
 & Size of .wasm files &  &  & ~2.9e-03 \\
 & Size of .scss files &  &  &  \\
 & Size of .less files &  &  &  \\
 & Size of .drconf files &  & ~1.2e-03 & ~5.3e-03 \\
 & Size of .url files &  &  &  \\
 & Size of .log files &  &  &  \\
 & Size of .wav files &  &  &  \\ 
\hline
\multirow{7}{*}{Privacy Practice} & Location &  & -0.88e-03 & -8.5e-03 \\
 & User Activity &  & ~4.3e-03 & -5.7e-03 \\
 & Website Content &  & ~3.2e-03 & ~8.5e-03 \\
 & Web History &  & -0.75e-03 &  \\
 & P.I.I. & -0.20e-03 & -2.7e-03 & ~0.88e-03 \\
 & Authentication Information & -0.078e-03 & ~4.5e-03 & ~8.7e-03 \\
 & Personal Communications &  & ~5.1e-03 & ~4.6e-03 \\ 
\hline
\multirow{3}{*}{\begin{tabular}[c]{@{}l@{}}User\\Perspective\end{tabular}} & \# of Users &  &  &  \\
 & \# of Raters &  &  &  \\
 & Rating Score &  & ~0.11e-03 & -4.2e-03 \\ 
\hline
\multirow{4}{*}{\begin{tabular}[c]{@{}l@{}}Activation\\Mode\end{tabular}} & Logged in & -0.32e-03 & -1.8e-03 & -4.0e-03 \\
 & Access to webpage granted &  & -0.61e-03 & -4.2e-03 \\
 & Non-designated mode &  & ~5.2e-03 & ~6.5e-03 \\
 & Fully-inactive mode & ~0.75e-03 & ~27e-03 &  \\
\hline
\end{tabular}
}
% \begin{tablenotes}
%       \small
%       \item All values are on the order of e-03.\heng{it would be better to put e-03 directly in the table, as it's very easy miss the note here.}
% \end{tablenotes}
% \vspace{-6mm}
\end{table}
\noindent\textbf{Analyzing the influences of the factors.}
% \bihui{As each factor more or less impacts on the browser performance, we predominately consider the most significant factors (i.e., the top 20\% determined by the {\tt varImp} function) that account for 80\% of the total significant factors, based on the Pareto principle (i.e., 80/20 rule), as proposed by \citet{Pareto}.}
% The Wald Chi-Square statistic is used to assess the statistical significance of a particular variable (e.g., extension size) on the model fit. %, as determined by the degrees of freedom.
% The Wald test examines the significance of a particular
% variable against the null hypothesis that the corresponding coefficient is equal to zero~\citep{Chi2}. 
% Subsequently, the resulting Wald statistic can be compared against a Chi-Square distribution to get a p-value that indicates the significance of the coefficient.
% The significant variables have a Pr(Chisq) (i.e., the p-value associated with the Chi-Squared statistic) less than 0.05. 
% Specifically, we use the {\tt Anova} function from the {\tt car} library in R \citep{Anova} to perform the Wald Chi-Square test.
% \heng{You only mentioned direction; also mention significance: how you got significant factors}
Features with a non-zero coefficients are significant and relevant for the model, as the non-significant ones are already pruned by the lasso penalty.
We determine the effect direction of a significant factor to indicate whether a significant factor has a positive or negative impact on a performance metric.
The effect direction of a significant factor is in fact the sign ($+$ or $-$) of the coefficient of the significant factor in the elastic net regression models.

% % what is signif. and effect?
% % what does the positive relationship represent for?
% \textbf{Interpretation of variables in the results.} Asterisks in the output of the models, presented in Tables \ref{tab:lmer-time}, \ref{tab:lmer-payload-energy}, and \ref{tab:lmer-stabilized-energy}, indicate the significant variables through the ANOVA (Analysis of variance) test. 
% The significant codes (signifi. codes), represented by asterisks or a decimal point, indicate the level of certainty that the coefficients have an impact on the dependent variable.
% For instance, ** represents that p-value is less than or equal to 0.01, meaning there is less than a 1.00\% chance that the coefficients might be insignificant.
% These significant variables have a Pr(Chisq) less than 0.05, where Pr(chisq) is the p-value that is associated with the Chi-Squared statistic. 
% The values of the Chi-Squared test demonstrate whether the presence of a particular independent variable leads to a statistically significant difference in the model, as determined by the degrees of freedom. 
% The effect column shows the the relationship of how the measured attributes of the extensions are affected as the variables change.
% Variables that have a positive correlation with the measured attribute of the extension are indicated by upward arrows, while the downward arrows indicate variables with a negative relationship.

\noindent\textbf{\underline{Findings:}}
% We present our finding in two parts: 1) External characteristics of the extensions (e.g., the extension size and extensions' activation modes), and 2) The impact of privacy practices on the measured attributes.
% \heng{Merge the the first three findings as they have been discussed in RQ2. Leading sentence: Our model analysis confirms that the activation modes of extensions significantly impact the browser performance. }
\textbf{Our model analysis comports with the observations in RQ2, confirming that the activation modes of extensions significantly affect the browser performance.}
Logged-in extensions tend to have a faster page load time and lower energy consumption. The negative relationships between \textit{Logged in} and the performance metrics suggest that logging in to the extension can improve browser performance. 
% Recalling the observations in RQ2, we can see that the page load energy consumption of the extensions without login is lower\heng{higher?} than it is in the fully-loaded mode \heng{no precise, as it does not apply to all tested extensions. maybe remove this sentence?}.
Therefore, it is recommended that users log in to the extension when prompted to do so.
Besides, it is advisable for users to grant permission to the extension to access the webpage, as indicated by the negative correlation shown in Table \ref{tab:lmer-stabilized-energy} - Activation Mode.
The negative correlation suggests that granting permission  (i.e., \textit{Access to webpage granted}) leads to an improvement in energy consumption.
% \textbf{The significance of the factor {\tt isInactive} compared to other activation modes implies that the improper use of the extensions, i.e., not using them on the designated website, can lead to a deterioration in the browser performance.}
The significance of the factor \textit{non-designated mode} and its positive correlation with the energy consumption metrics imply that improper use of extensions can negatively impact browser performance on energy consumption.
Similarly, it is observed that the \textit{fully-inactive mode} significantly jeopardizes the browser performance.
%When the use of the extension is ignored, only the page load time and the page load energy consumption are affected, without notable impact on the stabilized energy consumption.
 % Using an extension on a non-designated website can significantly impact payload performance, only affecting response time and energy consumption without affecting stabilized energy consumption.
%\bihui{Moreover, additional resources required to run the extensions in the non-designated and fully-inactive modes, can lead to the deterioration of browser performance.}
\textbf{In summary, adhering to proper usage scenarios, i.e., logging in to the extension, granting access to websites, and utilizing extensions on the designated websites, can benefit the overall browser performance.}

% \heng{I suggest moving this part to be the second or the first point in the results (ranking the order of the results based on their interestingness). }
%\textbf{The implementation of privacy practices within extensions can have varying impacts on browser performance.}
\textbf{Collection of user information by extensions (i.e., adoption of privacy practices) can lead to a significantly negative impact on energy consumption.}
%In particular, the adoption of privacy practices in the extensions significantly influences the browser performance in energy consumption.
For example, utilizing \textit{authentication information} within an extension is associated with an increase in the stabilized and page load energy consumption. 
When \textit{authentication information} is retrieved, it often involves communication with external servers, ongoing data transfers, and periodic page alive detection, %.
%Such continuous activity and communication require constant network connectivity and data processing, 
leading to increased energy usage. 
The use of \textit{personal communications} is correlated with increased energy consumption.
%The channel of \textit{personal communications} involves lightweight data transfers and background processes. 
% Unlike \textit{authentication information} or \textit{user activity}, %that may require extensive network requests, data transfers, or complex computations,  
The \textit{personal communication} practice monitors information, such as emails, texts, and chat messages, which involves continuous background processes and monitoring that may require active network connections and processing power,
resulting in increased energy consumption.
Moreover, analyzing and processing personal communications data in real-time can also introduce additional computational overhead, consuming more energy resources.
\bj{To understand our findings, we post our results to the Chrome extension developer forum, one explains that the additional energy consumption related to \textit{authentication information} and \textit{personal communication} is attributed to the use of APIs and network communications by extensions. Most of the extensions utilize the same thread as that of the browser rather than employing a Worker to perform such tasks. This approach ultimately contributes to higher energy consumption and places an increased load on the browser.}
% Interactions with other factors may lead to a negative correlation with page load time, but the effect is ignorable compared to others.

\textbf{Using \textit{website contents} exhibits a positive correlation in both page load energy consumption and stabilized energy consumption, emphasizing that curtailing the use of \textit{website contents} could result in deteriorated browser performance.}
Website contents, such as images, scripts, and multimedia elements, can engage users to spend more time on the webpage and require ongoing network requests and data transfers, resulting in additional energy consumption.
During a page load, the positive correlation suggests that engaging content may contribute to the increased energy consumption, possibly due to the rendering of dynamic and interactive elements.
For stabilized energy consumption, the positive correlation indicates that the ongoing user engagement may involve periodic updates, animations, or dynamic content, which can contribute to sustained energy consumption. However, this may lead to a more satisfying user experience, justifying the positive correlation.
Moreover, monitoring \textit{website contents} could lead to increased stabilized energy consumption, as the system remains active to support ongoing interactions and content updates.

\textbf{While monitoring \textit{location} and \textit{user activity} in the background can be resource-intensive, consuming unnecessary energy, a negative correlation observed between monitoring \textit{location} and \textit{user activity} and the stabilized energy consumption suggests a potential avenue for enhancing browser performance.} 
The collection of \textit{location} and \textit{user activity} information occurs during the page load phase, as evidenced by the positive (i.e., 4.3e-03) and nearly positive (i.e., -0.88e-03) relationships between \textit{location} and \textit{user activity} and the page load energy consumption.
Stabilized energy consumption pertains to the period following the page load while the user interacts with the loaded content. 
% \heng{not clear how this mitigation is done}
By mitigating the energy-intensive background processes associated with location and user activity during the stabilized period, such as reducing the frequency of background processes during low user activity and intensifying it during active user engagement, the browser ensures energy efficiency.
Consequently, such actions can significantly improve the stabilized energy consumption, albeit at the cost of a potential negative impact on the page load energy consumption.

\textbf{A greater number of lines of code (LOC) does not necessarily result in higher energy consumption, but the augmented count of unique paths through the code (NPATH) does.}
In the Code Metrics group, various metrics are evaluated for their impact on energy consumption. 
Notably, the number of lines containing source code (LOC) demonstrates a substantial negative effect, as shown in Table \ref{tab:lmer-stabilized-energy} (Code Metric), suggesting that an increase in the number of LOC that make up extensions does not necessarily lead to a negative impact on energy consumption (i.e., both stabilized and page load energy consumption) but may exhibit the opposite effect.
% Table \ref{tab:lmer-stabilized-energy} (Code Metric) shows a negative relationship between LOC and stabilized energy consumption, indicating that an increase in lines of code that make up extensions does not necessarily lead to a negative impact on stabilized energy consumption but may surprisingly exhibit the opposite effect. \heng{The two previous sentences repeat each other. Also mention page load energy consumption (with a negative coefficient too).}
Plausible explanations are: a larger codebase allows for more efficient and optimized implementations. 
A well-structured and streamlined codebase, even if extensive, may have undergone optimization practices that enhance energy efficiency.
Additionally, a larger codebase indicates a mature and well-maintained extension. 
Developers who invest time in refining and optimizing their codebase may prioritize user experience as part of their development practices, which could lead to a more resource-efficient extension, resulting in lower stabilized energy consumption.

\textbf{The positive relationship between code complexity (i.e., NPATH) and the stabilized energy consumption suggests that increased code complexity, as indicated by a higher number of unique paths, is associated with higher stabilized energy consumption.}
One example of NPATH is shown in Figure \ref{fig:code}, and another example is when using the {\tt sscanf} function in C to read data from a string, where can result in n times reading (i.e., executions) through this single statement (i.e., NPATH = n) if the string subsumes n lines.
% \heng{the ``another example'' is very vague}.
Developers are advised to consider strategies, such as adopting switch statements or hash maps to insert  projects to reduce code complexity and steering away from approaches like {\tt sscanf} that may contribute to elevated stabilized energy consumption. For example, developers can use the {\tt hook} function rather than {\tt sscanf} to read data when needed.

\textbf{The number of functions is impactful on stabilized energy consumption, indicating a strong positive relationship that an increase in the number of functions in extensions positively correlates with degraded browser performance in terms of stabilized energy consumption.}
An  elevated count of functions in extensions typically leads to heightened overhead execution, including an increase in callbacks and event handling, as well as the introduction of complex interdependencies.
% \heng{Probably wrong: positive coefficient means a higher number of functions is bad for performance. Fix the following as well.} 
This complexity can give rise to frequent interactions and updates, causing the browser to grapple with the management and coordination of various functions. 
Thus, suboptimal performance during the stabilized phase may ensue that in turn lead to increased energy consumption during the stabilized phase.
Additionally, an increase in the number of functions can lead to resource fragmentation. 
In this context, both the browser and the extension may engage in frequent allocation and deallocation of resources.
Due to the unavailability of certain resources, one or more of the available resources remain underutilized or unutilized \citep{functions}. 
This resource fragmentation introduces inefficiencies in resource management, thereby adversely affecting energy consumption during the stabilized phase.
% Modularity allows for better organization and isolation of functionalities, enabling the browser to execute specific tasks without unnecessary overhead. 
% This modular structure can contribute to efficient resource management, positively impacting stabilized energy consumption after the webpage has loaded.
% In stabilized energy consumption, the browser has already loaded the webpage. 
% An increased number of functions may allow for more selective activation of specific functionalities based on user interactions, reducing unnecessary energy consumption. 
% The ability to activate only relevant functions in response to user actions can lead to a more optimized and energy-efficient browsing experience.

\begin{figure}[ht]
%\vspace{-4mm}
\caption{A NPATH example in python.}
%\vspace{-2mm}
\label{fig:code}
\begin{lstlisting}
# NPATH = 4: (a,b) - (1,1), (1,0), (0,1), (0,0)
# where (1,1) means functions go_a and go_b are called
# Each if-statement yeilds two paths (i.e., True - call the function and False - do nothing and go to the next).
def npathDemo (a, b):
    if a:
        go_a()
    if b:
        go_b()
\end{lstlisting}
%\vspace{-6mm}
\end{figure}

\textbf{The size of various file types within the source files of extensions exhibit diverse effects on energy consumption.}
For instance, the file characteristics outlined in Table \ref{tab:lmer-stabilized-energy} reveals a negative correlation between the size of .png and .eot typed files and energy consumption, implying that an increase in the utilization of image files (.png) and OpenType font files for webpages (.eot) does not result in heightened energy consumption; instead, it suggests a favorable impact.
In contrary to the .png typed files, .svg typed files, which are also image files, have a negative impact on stabilized energy consumption. 
SVG images consist of a set of instructions that require execution, while png files are composed of pixels that can be loaded more efficiently.
It is advisable for developers to take into consideration the impact of file types on browser performance.
For instance, it is recommended to substitute the use of .svg files with .png files and consider employing .mp3 or .wav files, which exhibit no significant correlation with browser performance, as opposed to .ogg audio files to enhance the overall browser performance.

\noindent\textbf{\underline{Summary and implication}:} 
% \heng{I suggest recording the summaries based on their interestingness: activation modes, privacy practices (data collection), and then others (the last part can be shortened if not quite conclusive).}
%Our findings indicate that the activation modes of extensions and the use of privacy practices are closely related to browser performance. We recommend that users adopt proper usage practices when using the extensions. 
Both the privacy practices and activation modes of extensions significantly impact browser performance.
%Our analysis confirms the impact of the activation modes on energy consumption: 
Adhering to proper usage practices, i.e., logging in to the extension, granting access, as well as utilizing the extension on the designated websites, benefits the overall browser performance.
Besides, it is crucial for extension developers and users to remain vigilant regarding the impact of privacy practices (i.e., collection of user data) on energy consumption when adopting or accepting the privacy practices. \bj{Extension developers are suggested to consider using  separate Worker threads to perform tasks rather than relying on the main browser thread to optimize the extensions.} 
In the context of source code, we recommend that developers focus on reducing code complexity, e.g., adopting hash maps, and minimizing the use of a high number of functions which help mitigate the associated rise in stabilized energy consumption.
Furthermore, when constructing extension projects, it is advisable to flexibly employ certain file types, such as PNG and EOT files, to optimize performance.
%Besides, we observe that extensions with larger code sizes tend to be more energy efficient.
%We also observe the significant impact of the privacy practices of extensions on browser performance. 
%Our findings provide insights for users to select performance-efficient extensions and for developers to optimize their development decisions.
% \end{tcolorbox}
% \vspace{-2mm}

\section{Threats to Validity}
\label{TV}
%The goal of this study is to discern the performance impact of browser extensions.
In this section, we discuss the threats to the validity of the study.

\textbf{Threats to Internal Validity.} %pertain to factors that undermine the accuracy and credibility of research findings by introducing bias or distorting the relationship between the variables under investigation.
\bj{In our study,  around 90\% of the total extensions, gathered in Section \ref{CE}, are removed. 
60\% of the total extensions without privacy practice specifications are discarded because of their untraceable structural information for later analysis. 
Given the vast number of extensions available, it becomes impractical to study each one individually. 
We opt to cluster the rest of extensions and  select the representative one from each category to stand in for the entire group. 
We want to ensure a diverse representations across the 11 categories in our dataset.
}
\bihui{In addition, the sample of extensions studied may not be representative of the broader population of extensions, which could introduce sampling bias. 
To mitigate this, we keep the selection of extensions diverse reflecting different categories, popularity levels, the use of different privacy practices, and functionalities.}
In RQ3, we use elastic net regression model to study the relationship between the various factors of an extension and its performance impact. However, the correlation may not suggest causation.
To mitigate this threats, we have proposed 105 possible factors to consider as many factors as possible.
Nonetheless, there may still exist other confounding factors not considered in this study. 

\textbf{Threats to External Validity.} %refer to factors that challenge the generalizability of the findings of a research study to settings or circumstances beyond the study sample and context. 
% Nevertheless,
In this work, we only studied the Google Chrome browser. 
Our results may not be generalized to other browsers. 
Nevertheless, as Chrome is the most popular browser, our findings can benefit a large number of browser users and extension developers. 
In addition, popular browsers such as Chrome and Safari are using similar architectures. Our findings may provide similar insights for other browsers.
%We conduct our experiments on only one desktop computer system using the Linux operating system.
% platform running a specific operating system, i.e., Linux, which poses a threat to the external validity of the experiment.
%We may not claim the same energy or run-time performance implications on other computer systems or operating systems.
Compared to prior studies (e.g., 3 extensions tested on \citet{Pearce}’s work, 5 extensions tested on \citet{Merzdovnik}'s work, and 8 extensions tested on \citet{Borgolte}’s work), our extension numbers (72) outweigh theirs, and \bj{our extensions types are more comprehensive}.

\textbf{Threats to Construct Validity.}
%\heng{Maybe move this to be ``Construct Validity''?}
Threats to the construct validity of our study may involve network instability, running background processes and daemons, such as a routing daemon that handles multiple routing protocols and replaces routed, which may impact the stability of our measurements. %\heng{give an example? maybe also network instability?} while we collect the measurements.
It is challenging to achieve complete control over network stability and all background processes or daemons of a system. However, we try to reduce the impact of background processes and daemons as much as possible by stopping them while we collect the measurements.
\bihui{We assume a certain users' usage patterns when we use Selenium to simulate users' interactions with extensions in distinct activation modes of extension.
Nevertheless, individual user settings and preferences are unpredictable.
Despite our efforts to account for various common conditions and settings, our findings may not be generalizable to cover all the possible real-world user interactions with extensions.}

\section{Related Work} 
\label{RW}
In this section, we present prior work with respect to browser and extension performance. %the energy consumption and page load time of Google Chrome extensions.
\vspace{-6mm}

\subsection{Browser Performance} \label{relatedwork2}
% \bihui{background}
 Studies, such as \citet{Macedo, Janssen, Tian}, investigate the run-time performance and the energy consumption of browsers and web applications.
\textit{\citet{Macedo}} compare the energy consumption of Google Chrome and Mozilla's Firefox when browsing webpages and find that Google Chrome is more energy-efficient when navigating web pages, but uses more energy for RAM, particularly when interacting with YouTube.
\textit{\citet{Janssen}} conduct a study to investigate the effect of the Critical CSS technique on the run-time performance and the energy consumption of Android mobile web applications in Google Chrome and Mozilla Firefox. \textit{Janssen et al.} find that the technique can positively improve the run-time performance of Android mobile web apps slightly, but it has no significant impact on energy consumption.
\textit{\citet{Tian}} analyze the quality of user experiences across three mobile browsers - Chrome, Firefox, and Opera - by collecting data from 337 webpages and analyzing the loading time and cache performance. \textit{Tian et al.} show that a significant proportion of webpages exhibit notable variations in terms of loading time and cache performance across different browsers.

Prior work \citep{Janssen, Tian, Macedo} focuses on examining the browsing performance across various browsers but does not explore the impact of browser extensions on browser performance. Our work, in contrast, studies the impact of extensions on browser performance. %encompasses a broader scope as we not only evaluate the runtime performance, such as page loading time and the page load energy consumption but also analyze the energy changes after page stabilization.
% As such, their work is complementary to ours. 
%As such, this makes our study distinct and complementary to existing work.

\subsection{Extension Performance}
% \bihui{background}
% In comparison to the studies in the previous subsections \ref{relatedwork1} and \ref{relatedwork2}, 
In comparison to the studies discussed in \ref{relatedwork2}, prior studies, such as \citet{Pearce, Merzdovnik, Borgolte}, delve deeper into the effect of privacy-focused browser extensions on browser performance.
\textit{\citet{Pearce}} explores the potential of three open source advertisement (ad) blockers to reduce the page loading time by eliminating ads from internet browsing and video streaming. 
The evidence indicates that ad blockers are effective in saving energy due to their ability to shorten page loading time.
%\textit{Pearce}'s work is complementary to our work, where we encompass a broader range of extension types and provide an in-depth analysis, i.e., considering multiple performance metrics, different activation modes, and various performance-influencing factors, while \textit{Pearce} focuses solely on the browsing quality \heng{what is browsing quality of experience? fix this -\bihui{revised}} with ad-blocking extensions. %, our study encompasses a broader scope.
\textit{\citet{Merzdovnik}} evaluate the effectiveness and the system performance impact of 5 anti-tracking extensions (e.g., Ad-Block Plus) across 100,000 websites.
The results show that these anti-tracking extensions do not increase CPU time, however, they consume more memory. The metrics used in the study, such as memory consumption, are comparable to the performance metrics used in our study. 
%Our study places a greater emphasis on the impact of the extensions on browser performance \heng{they also look at performance?} and delves deeper into the underlying factors that cause these impacts.
\textit{\citet{Borgolte}} analyze the impact of eight privacy-focused browser extensions (e.g., Ad-Block Plus and Privacy Badger) on user experience and system performance in both Google Chrome and Mozilla Firefox.
Overall, the results indicate that these privacy-conscious extensions do not impede the system performance and even improve the user's browsing experience.

Prior studies \citep{Pearce,Merzdovnik,Borgolte} predominantly focus on the examination of a single type of extensions, particularly those associated with activity blocking.
Our study, however, stands out by encompassing representative extensions from 11 diverse categories, which cover various functional types of extensions.
In addition, our study considers a different set of performance metrics (e.g., including energy consumption) and delves deeper to understand how  different activation modes and other factors of the extensions (e.g., privacy practices) influence the browser performance. 
\bihui{Although our study focuses on a smaller set of websites, 
% which is a quid pro quo trade-off for our approach, it is important to note that
the websites used in our experiments are highly representative, as they are selected from 11 categories in Google Chrome web store and commonly utilized by users worldwide.  
Therefore, our findings still hold the relevance and provide valuable insights into the performance of different extension types across widely-used online platforms.}
%These extensions are categorized based on their characteristics, such as the use of privacy practices, and we assess the performance impact of various activation modes of extensions utilizing more comprehensive metrics that concentrate specifically on the browser performance. 
%Furthermore, we go beyond simply measuring and comparing energy consumption and delve into the underlying factors that lead to spikes in energy and page load time.
% we not only measure and compare energy consumption, but we also investigate the factors that contribute to peaks in energy and response time of payload.

\section{Conclusion}
\label{CON}
In this paper, we study the impact of extensions on browser performance, specifically in terms of page load time and energy consumption.
We observe that browser performance can be negatively impacted by the use of extensions, even when the extensions are used in unexpected circumstances (e.g., not logged in or access to a webpage not granted) or are not active (e.g., not used for designated websites or fully deactivated).
We also observe that the privacy practices of an extension are significantly correlated with its performance impact. % and that extensions with larger code sizes tend to be more energy efficient.
Our work provides the following recommendations for extension users and extension or browser developers:

\begin{itemize}[leftmargin=*,topsep=0pt]

\item Browser and extension developers should be vigilant about the performance impact of browser extensions, particularly in regard to energy efficiency. For example, they could provide warnings and information about the performance impact of the extensions or suggest users to disable the extensions in performance-critical scenarios. %, and optimize the extensions or browser-extension integration in terms of performance.

\item Extension developers and users should be aware of the performance impact of different activation modes of extensions. Users are suggested to adhere to the proper usage practices of extensions (e.g., granting login to an extension when required), while developers are suggested to spend more effort on the performance testing and optimization of the intended usage scenarios of extensions. 
%Third, browser developers are suggested to develop mechanisms to reduce the performance impact of extensions when they are not providing proper functionalities (e.g., when in a non-designated mode) to users.
%Finally, our findings also provide insights for users to select performance-efficient extensions and for developers to optimize their development decisions. 

\item When constructing extension projects for optimal performance, extension developers should be mindful of the influence of code complexity and the usage of certain file types. 
\bj{For example, reducing code complexity (e.g.,
adopting hash maps) and minimizing the use of a high number of functions are helpful to mitigate the associated rise in stabilized energy consumption. 
Moreover, it is advisable to flexibly employ certain file types, such as using PNG files in place of  SVG, and using EOT files in place of other font files, to optimize performance.}

\item Extension developers and users should pay attention to the potential performance impact associated with privacy practices when adopting or accepting privacy practices for an extension. 
\bj{Extension developers are advised to optimize extensions by utilizing a dedicated Worker thread for tasks instead of relying solely on the primary browser thread. 
Extension users, in turn, should be aware that privacy practices (e.g., \textit{website contents} and \textit{personal communication}) could drain more energy consumption, and this consideration should influence their choices when selecting extensions.}
\end{itemize}

In the future, we will extend our dataset of extensions to conduct experiments on more extensions. 
Furthermore, future work will explore additional factors (e.g., programming language used) that may influence extension performance.

% \section{Data Availability}
% \label{DA}
% % Our replication package can be found at a GitHub repository~\citep{RepPackage}.
% We share our replication package \citep{RepPackage2} on \url{https://github.com/Bihui-Jin/suppmaterial-impact-of-extensions-on-browser-performance}
%  for future work to build on our work.

\printbibliography

\newpage
\appendix
\section{Appendix A: The number of downloads of selected 71 extensions}\label{Appendix1}
\begin{center}
\setlength\LTleft{-14mm}
{\small\tabcolsep=3pt
\renewcommand{\thetable}{A.1}
\begin{longtable}{lrrr}
% \begin{table}
% \centering
% \renewcommand{\thetable}{A.1}
\caption{Download details for 72 selected extensions.}
\label{tab:downloads}\\
% \resizebox{\linewidth}{!}{%
% \begin{tabular}{lrrr} 

\toprule
\multicolumn{1}{c}{}
& \multicolumn{1}{r}{Extension Name} & \multicolumn{1}{c}{Extension ID} & \multicolumn{1}{c}{Number of downloads}  \\ \hline
\endfirsthead

\multicolumn{4}{c}%
{{ \tablename\ \thetable{} -- continued from previous page}} \\
\hline 
\multicolumn{1}{c}{}
& \multicolumn{1}{r}{Extension Name} & \multicolumn{1}{c}{Extension ID} & \multicolumn{1}{c}{Number of downloads} \\ \hline 
\endhead

\hline \multicolumn{4}{|r|}{{Continued on next page}} \\ \hline
\endfoot

\hline \hline
\endlastfoot
% \toprule
%  & \multicolumn{1}{c}{Extension Name} & \multicolumn{1}{c}{Extension ID} & \multicolumn{1}{c}{Number of downloads} \\ 
% \midrule
1  &    LM Note Generator For ESPN Fantasy Football &  ahcblhpcealjpkmndgmkdnebbjakicno &       1,000+ \\
2  &                                   Picwatermark &  aiiimepjikpdipbpmknolbnjbeohbmaa &           22 \\
3  &                          Dark Mode - Night Eye &  alncdjedloppbablonallfbkeiknmkdi &     200,000+ \\
4  &                                            7TV &  ammjkodgmmoknidbanneddgankgfejfh &     900,000+ \\
5  &              Neeva Search + Protect for Chrome &  aookogakccicaoigoofnnmeclkignpdk &      20,000+ \\
6  &                      Ultimate Video Translator &  bboamecjefgpaemgfpcjeediamdnkklc &      10,000+ \\
7  &      Better Tab: Speed Dial, News Feed \& To-do &  behkgahlidmeemjefcbgieigiejiglpc &          162 \\
8  &                            Send to Google Maps &  bhggankplfegmjjngfmhfajedmiikolo &      30,000+ \\
9  &                    Octotree - GitHub code tree &  bkhaagjahfmjljalopjnoealnfndnagc &     400,000+ \\
10 &        Truffle.TV (formerly known as Mogul.TV) &  bkkjeefjfjcfdfifddmkdmcpmaakmelp &     100,000+ \\
11 &              Tab Resize - split screen layouts &  bkpenclhmiealbebdopglffmfdiilejc &     700,000+ \\
12 &  LINER - Search Faster \& Highlight Web/Youtube &  bmhcbmnbenmcecpmpepghooflbehcack &     400,000+ \\
13 &                                   Watch2Gether &  cimpffimgeipdhnhjohpbehjkcdpjolg &     900,000+ \\
14 &                           Adblock for Youtube™ &  cmedhionkhpnakcndndgjdbohmhepckk &  10,000,000+ \\
15 &                               Image Downloader &  cnpniohnfphhjihaiiggeabnkjhpaldj &   1,000,000+ \\
16 &                  BuiltWith Technology Profiler &  dapjbgnjinbpoindlpdmhochffioedbn &     300,000+ \\
17 &                                    Sourcegraph &  dgjhfomjieaadpoljlnidmbgkdffpack &     100,000+ \\
18 &                                  AmazingHiring &  didkfdopbffjkpolefhpcjkohcpalicd &      20,000+ \\
19 &   Ecosia - The search engine that plants trees &  eedlgdlajadkbbjoobobefphmfkcchfk &   2,000,000+ \\
20 &                                    Dark Reader &  eimadpbcbfnmbkopoojfekhnkhdbieeh &   4,000,000+ \\
21 &                                   FrankerFaceZ &  fadndhdgpmmaapbmfcknlfgcflmmmieb &   1,000,000+ \\
22 &                              Microsoft Rewards &  fbgcedjacmlbgleddnoacbnijgmiolem &   2,000,000+ \\
23 &          GoFullPage - Full Page Screen Capture &  fdpohaocaechififmbbbbbknoalclacl &   5,000,000+ \\
24 &      Trusted Shops extension for Google Chrome &  felcpnemckonbbmnoakbjgjkgokkbaeo &     300,000+ \\
25 &                                 webpage cloner &  ffjnfifmelbmglnajefiipdeejghkkjg &            7 \\
26 &                        Tumblr – Post to Tumblr &  ffnhmkgpdmkajhomnckhabkfeakhcamm &       4,000+ \\
27 &                              Elfster's Elf It! &  fhjanlpjlfhhbhbnjohflphmfccbhmoi &      10,000+ \\
28 &                 Dynatrace Real User Monitoring &  fklgmciohehgadlafhljjhgdojfjihhk &     400,000+ \\
29 &               Use Immersive Reader on Websites &  fmidkjgknpkbmninbmklhcgaalfalbdh &     100,000+ \\
30 &                                  Tricky Enough &  fnhmjceoafkkibpijbfpfajbhkknadmb &            0 \\
31 &           FantasyPros: Win your Fantasy League &  gfbepnlhpkbgbkcebjnfhgjckibfdfkc &     200,000+ \\
32 &               SimplyCodes $|$ Coupons that work. &  gfkpklgmocbcbdabfellcnikamdaeajd &      10,000+ \\
33 &                                        NekoCap &  gmopgnhbhiniibbiilmbjilcmgaocokj &       1,000+ \\
34 &                          Pinterest Save button &  gpdjojdkbbmdfjfahjcgigfpmkopogic &   7,000,000+ \\
35 &               Wappalyzer - Technology profiler &  gppongmhjkpfnbhagpmjfkannfbllamg &   1,000,000+ \\
36 &                                    coffeelings &  hcbddpppkcnfjifbcfnhmelpemdoepkk &     200,000+ \\
37 &                       Stem Player Album Upload &  iedjpcecgmldlnkbojiocmdaedhepbpn &          181 \\
38 &                                  Boxel Rebound &  iginnfkhmmfhlkagcmpgofnjhanpmklb &   1,000,000+ \\
39 &                                          Ampie &  ikdgincnppajmpmnhfheflannaiapmlm &          414 \\
40 &                                        Weather &  iolcbmjhmpdheggkocibajddahbeiglb &     100,000+ \\
41 &                           Enablement Assistant &  jbebkmmlkhioeagiekpopmeecaepaihd &     400,000+ \\
42 &                                        OkTools &  jicldjademmddamblmdllfneeaeeclik &     100,000+ \\
43 &                                 Designer Tools &  jiiidpmjdakhbgkbdchmhmnfbdebfnhp &      30,000+ \\
44 &                              ER-help Extension &  jpefkkpmalfnilnbghfnjodceifpemdb &       3,000+ \\
45 &                                 Automation 360 &  kammdlphdfejlopponbapgpbgakimokm &     100,000+ \\
46 &            RSS Reader Extension (by Inoreader) &  kfimphpokifbjgmjflanmfeppcjimgah &      40,000+ \\
47 &                                      Inforness &  kgaebnfbgpcnglnhjhglinfiecgccfij &            5 \\
48 &                                NFL Live Scores &  kimjfkgkpmafgngclkdpjdlkdlghoikh &       1,000+ \\
49 &  \begin{CJK}{UTF8}{mj} Naver/Daum Media Filter(네이버/다음 뉴스 언론사 표시/차단) \end{CJK} &  kpghljlpdknmomchobaoecdlkcpocaga &       7,000+ \\
50 &                          Bulk Image Downloader &  lamfengpphafgjdgacmmnpakdphmjlji &      30,000+ \\
51 &                                 imgur Uploader &  lcpkicdemehhmkjolekhlglljnkggfcf &      10,000+ \\
52 &                                   JobsAlert.pk &  ldjnabbinoccbodkejkdiolmadimbjkj &           28 \\
53 &                          Lichess Opponent Form &  lipplpkgbnhdfdchoibgafjdblpjdkpi &           36 \\
54 &                                 A dónde Viajar &  lnphplhkejidgcncalbkbngbiafmjnml &           32 \\
55 &                RotoGrinders - DraftKings Tools &  lokmacldfjfgajcebibmmfohacnikhhd &      10,000+ \\
56 &                                        Feedbro &  mefgmmbdailogpfhfblcnnjfmnpnmdfa &      40,000+ \\
57 &                                      TubeBuddy &  mhkhmbddkmdggbhaaaodilponhnccicb &   1,000,000+ \\
58 &         Aerobi - Enhance Your YouTube Workouts &  mlfkmhibffpoleieiomjkekmjipdekhg &           16 \\
59 &          Screencastify - Screen Video Recorder &  mmeijimgabbpbgpdklnllpncmdofkcpn &   6,000,000+ \\
60 &                                WAM: WordSeeker &  mpejojclnbakefnlfmnkaaianojbicdk &          158 \\
61 &               CiteMaker CiteWeb $|$ APA 7th Edn. &  naankklphfojljboaokgfbheobbgenka &       4,000+ \\
62 &                   Keepa - Amazon Price Tracker &  neebplgakaahbhdphmkckjjcegoiijjo &   2,000,000+ \\
63 &                                       MetaMask &  nkbihfbeogaeaoehlefnkodbefgpgknn &  10,000,000+ \\
64 &              Bitwarden - Free Password Manager &  nngceckbapebfimnlniiiahkandclblb &   2,000,000+ \\
65 &                         ShadowPay Trademanager &  obhadkdgdffnnbdfpigjklinjhbkinfh &      70,000+ \\
66 &                      Custom Cursor for Chrome™ &  ogdlpmhglpejoiomcodnpjnfgcpmgale &   5,000,000+ \\
67 &                    Amazon Assistant for Chrome &  pbjikboenpfhbbejgkoklgkhjpfogcam &   8,000,000+ \\
68 &                                InteractiveFics &  pcpjpdomcbnlkbghmchnjgeejpdlonli &     100,000+ \\
69 &                              The Newsroom Beta &  pgfokhpgehbmeifbpdhegfnpaahabfja &          199 \\
70 &  \begin{CJK}{UTF8}{min} 買い物ポケット \end{CJK} &  pgmbeccjfkdbpdjfoldaahpfamjjafma &     500,000+ \\
71 &                           Global Twitch Emotes &  pgniedifoejifjkndekolimjeclnokkb &     100,000+ \\
72 &  Free Best VPN PC-Chrome-Unlimited Proxy Guide &  pkihbahhbihfoebgdfkibnblbhjfgefc &          207 \\
\bottomrule
\end{longtable}
}
\vspace{-2mm}
\begin{tablenotes}
    \normalsize
     \item Extensions can be visited at the Chrome Web Store via the website https://chrome.google.com/webstore/detail/\{ExtensionID\}
\end{tablenotes}
\end{center}

\end{document}